\documentclass[preprint,5p,times,twocolumn]{elsarticle}
\usepackage{amsmath,amsfonts}
\usepackage{algorithmic}
\usepackage{array}
\usepackage{textcomp}
\usepackage{url}
\usepackage{verbatim}
\usepackage[nohyperlinks]{acronym}
\usepackage{graphicx}
\usepackage{booktabs}
\usepackage{hyperref}
\usepackage{wrapfig}
\usepackage[most]{tcolorbox}

\newtcolorbox{mytextbox}[1][]{%
  sharp corners,
  enhanced,
  colback=white,
  attach title to upper,
  #1
}

\journal{Pre-print}

\begin{document}
\begin{frontmatter}
\title{Silence Speaks Volumes:\\A New Paradigm for Covert Communication via History Timing Patterns}
\author[1]{Christoph Weissenborn}\ead{christoph.weissenborn@posteo.de}
\author[2]{Steffen Wendzel}\ead{steffen.wendzel@uni-ulm.de}
\address[1]{Federal Office for Information Security, Bonn, NRW, Germany}
\address[2]{Institute of Information Resource Management (IRM), Ulm University, Ulm, BW, Germany}

\begin{abstract}
A Covert Channel (CC) exploits legitimate communication mechanisms to stealthily transmit information, often bypassing traditional security controls. Among these, a novel paradigm called History Covert Channels (HCC) leverages past network events as reference points to embed covert messages. Unlike traditional timing- or storage-based CCs, which directly manipulate traffic patterns or packet contents, HCCs minimize detectability by encoding information through small pointers to historical data. This approach enables them to amplify the size of transmitted covert data by referring to more bits than are actually embedded. Recent research has explored the feasibility of such methods, demonstrating their potential to evade detection by repurposing naturally occurring network behaviors as a covert transmission medium.

This paper introduces a novel method for establishing and maintaining covert communication links using relative pointers to network timing patterns, which minimizes the reliance of the HCC on centralized timekeeping and reduces the likelihood of being detected by standard network monitoring tools. We also explore the tailoring of HCCs to optimize their robustness and undetectability characteristics. Our experiments reveal a better bitrate compared to previous work.
\end{abstract}

\begin{keyword}
Covert channels; Information hiding; Steganography; Information security; Censorship Circumvention
\end{keyword}

\end{frontmatter}
\acresetall

\begin{mytextbox}[colupper=blue,fontupper=\bfseries\normalsize]
This is a pre-print.\\The final and enhanced version of this paper was published under a different title in the \emph{Journal of Information Security and Applications (JISA)}: \url{https://doi.org/10.1016/j.jisa.2026.104431}
\end{mytextbox}

\section*{Acronyms}
\begin{acronym}[1200pt]
\setlength{\itemsep}{-3pt}
\acro{ARP}{Address Resolution Protocol}
\acro{CAF}{Covert Amplification Factor}
\acro{CC}{Covert Channel}
\acro{CR}{Covert Receiver}
\acro{CS}{Covert Sender}
\acro{DYST}{Did You See That}
\acro{ECC}{Error Correction Code}
\acro{FHSS}{Frequency-Hopping Spread Spectrum}
\acro{HCC}{History Covert Channel}
\acro{ICD}{Intra-Connection Delay}
\acro{ICPN}{Intra-Connection PDU Number}
\acro{IP}{Internet Protocol}
\acro{IPD}{Inter-Packet Delay}
\acro{IPT}{Inter-Packet Time}
\acro{ISD}{Inter-Signal Delay}
\acro{ISPN}{Inter-Signal PDU Number}
\acro{LAN}{Local Area Network}
\acro{NTP}{Network Time Protocol}
\acro{OOOD}{Out-of-Order Delivery}
\acro{PDU}{Protocol Data Unit}
\acro{POI}{Packets of Interest}
\acro{ROC}{Receiver Operating Characteristic}
\acro{SGA}{Simple Genetic Algorithm}
\acro{SHP}{Silent History Protocol}
\acro{TCP}{Transmission Control Protocol}
\acro{WAN}{Wide Area Network}
\end{acronym}

\section{Introduction}
The landscape of network cybersecurity is constantly evolving, marked by the perpetual back and forth between attackers seeking covert communication and defenders striving to lock down network environments~\cite{NIHBook}. A persistent challenge in cybersecurity is the utilization and detection of \ac{CC}s, which circumvent security policies by embedding hidden communications within legitimate network operations~\cite{petitcolas1999information,Schmid23,zander2007survey}. \ac{CC} may be used for malware command and control or data exfiltration~\cite{NIHBook,caviglione2022never,zander2007survey}, but also by journalists looking to report under surveillance \cite{wustrow2011telex,fifield2012evading} and citizens aiming to bypass Internet censorship \cite{AsiaCCS24:OPPR}. The primary impact of a \ac{CC} has been traditionally measured by the channel's number of bits transferred per second (bandwidth)~\cite{twenty02}.

\ac{CC} can be understood as steganographic methods that modify overt information objects using different hiding patterns to hide secret information from plain view~\cite{NIHBook,zander2007survey}. Traditional covert channels often rely on the modification of network traffic patterns or the timing of packet transmissions to embed hidden messages \cite{mileva14,wendzelpatternsurvey15}. However, these methods have been regularly investigated during the last 20 years and are now increasingly susceptible to detection by traffic analysis tools and intrusion detection systems, using methods such as anomaly detection~\cite{detect22}.

A recent advancement in this domain is the introduction of the \ac{HCC}~\cite{DYST22}. They minimize the detectable footprint of covert communication by referencing legitimate network traffic through small pointers rather than creating or modifying cover traffic. This approach enables an amplification of the covert message to be transferred (transferring a small pointer instead of the larger bitstring being pointed to). One such protocol suggested by Wendzel et al.\ is \textit{\ac{DYST}}~\cite{DYST22}. While it is protocol-agnostic in principle, its implementation leverages ARP requests to reference hashes of overt network traffic. Despite its innovative approach, \ac{DYST} is not without limitations, particularly in terms of detection risk due to signaling based on overt traffic, its dependence on clock synchronization, and scalability in high packet rate environments.

In response to these challenges, we propose the \ac{SHP}, a flexible approach that further leverages the history concept introduced by \ac{DYST} but addresses its key limitations. \ac{SHP} can use various network timing patterns to establish a \ac{CC} and tailor it to the network environment. This enhancement not only widens the applicability of history-based \ac{CC}s, but also mitigates issues related to clock synchronization and scalability, particularly in high-traffic environments.

The remainder of this paper is structured as follows. Sect.~2 provides a brief introduction to \ac{HCC} and covers related work. We discuss our approach in Sect.~3 and our implementation in Sect.~4. We conduct an in-depth evaluation in Sect.~5 and discuss our results in Sect.~6. Finally, Sect.~7 concludes and suggests future work.

\section{Background and Related Work}
\ac{CC}s are hidden communication paths that allow the transfer of information in a manner that circumvents system security policies~\cite{wendzelpatternsurvey15}. Traditionally, network-based covert channels have embedded secret messages in network communications by manipulating packet sizes, timing, or other traffic characteristics. Since the introduction of \ac{CC} by Lampson in 1973~\cite{lamp73}, researchers have focused on two primary methods~\cite{DYST22}: (1) active sending, where the covert sender creates its own traffic to embed secret data, and (2) passive sending, where legitimate traffic is modified to carry covert information. Both methods have a common limitation — they render the covert communication detectable due to the unnatural patterns they introduce into the network traffic. 

Traditional network \ac{CC}s embed information within the timing, packet sequence, or other characteristics of transmissions, and several methods are difficult to detect using conventional security measures~\cite{mileva14,wendzelpatternsurvey15,Iglesias2017Are,wang2016dwt,NIHBook,epskappa24}.

Network \ac{CC}s based on encoding data in timing information include \ac{IPD}-based ones such as Jitterbug~\cite{jitterbug16}, which use the time difference between packets as input, and packet-sorting-based encoding, which utilizes the order of a particular set of packets as a channel \cite{wendzelpatternsurvey15,cabuk04}. These approaches can also be combined into more complex encoding methods.

All of them may use the timing information of all packets from a given connection, or focus on specific \ac{POI}, such as \ac{ARP} requests, DNS queries, or other common protocol messages \cite{wendzelpatternsurvey15}.  \ac{POI}s may be chosen based on their attributes--such as length, timing, or classification--to optimize throughput, preserve legitimate network functionality, or avoid detection.

Several approaches have aimed to reduce attention to covert transmissions \cite{NIHBook,wendzelLowAttention}. A recent development led to \ac{CC}s that are capable of transmitting high-bandwidth data over encrypted channels by taking over and controlling encrypted overt communication~\cite{cckex24}. But in some cases even a \ac{CC} in encrypted traffic can be detected using machine learning on statistics such as the source address and the entropy of the timestamp~\cite{ai24,morepackets21,Ondov2022Covert,Epishkina2019Timing}. The likelihood of being able to discern overt and covert channels using such statistics increases as the ratio between covert and overt signals increases. This holds even for methods that specifically focus on detection in small samples~\cite{GAS22}.

\begin{figure}[h]
  \centering
  \includegraphics[width=\linewidth]{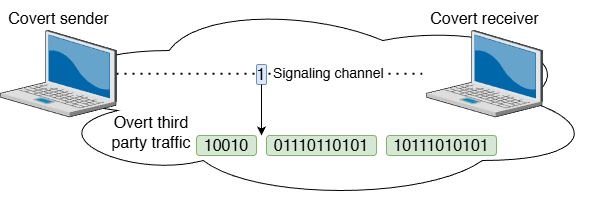}
  \caption{Illustration of a \ac{HCC}, where the sender references a secret message by embedding a small pointer to prior legitimate traffic.}
  \label{fig:hcc}
\end{figure}

History \ac{CC}s represent an advancement in this regard, because they communicate almost exclusively by \emph{pointing} to existing legitimate network traffic rather than altering or generating new traffic. This allows them to amplify the number of bits transferred in the channel by sending only a pointer covertly, as can be seen in Fig.~\ref{fig:hcc}. Amplification minimizes observable artifacts, making \ac{HCC} a promising avenue for evading modern detection systems while keeping the net amount of data transferred the same \cite{DYST22,wang5634454moss}. The ratio of bits effectively received compared to the number of bits used for the pointer is called the \ac{CAF}~\cite{DYST22}:

\[ CAF(algorithm) = \frac{bits_{message}}{bits_{pointer}}. \]

\begin{figure}[h]
  \centering
  \includegraphics[width=\linewidth]{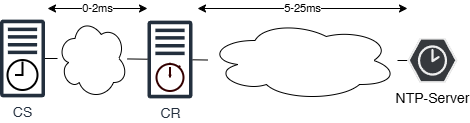}
  \caption{Clock synchronization inaccuracy for a \ac{CC} based on high accuracy clock synchronization. While \ac{CS} and \ac{CR} are close to each other, they may synchronize to an NTP server further away, thereby limiting the robustness of timing-based \ac{CC}.}
  \label{fig:clocksync}
\end{figure}

The pointer refers the receiver to unaltered legitimate traffic, whose hash value is then used as bits of the secret message. \ac{DYST} uses \ac{PDU} datafields like IP checksum as the input value in combination with the \ac{PDU}'s timestamp. This enables \ac{DYST} to use the potentially high entropy of timestamps for a data input source. However, this potential can only be realized in practice if the involved systems can reliably synchronize their clocks with a very high degree of accuracy -- ideally down to the same millisecond. \ac{DYST} relies on other protocols such as NTP for clock synchronization~\cite{DYST22}. Since prior studies suggest that NTP synchronization of internet systems can typically vary between tens of milliseconds~\cite{ntp13} and a few milliseconds~\cite{ntp19}, this need for clock synchronization limits \ac{DYST}'s effectiveness, even if the covert agents themselves are close to each other, as is visualized in Fig.~\ref{fig:clocksync}. If the covert agents are not overt systems themselves, their NTP synchronization would also be a suspicious signal in itself. Furthermore, clock drift or NTP errors can lead to desynchronization, compromising the effectiveness and reliability of the channel, which are essential for covert communication.

\noindent\textbf{Contributions.} This paper makes the following contributions:

\begin{itemize}

  \item We introduce \ac{SHP}, a history covert channel that uses relative pointers into naturally occurring timing patterns, eliminating the need for tight clock synchronization while preserving stealth by amplifying covert payload per signal.

  \item We provide a complete, reproducible parameterization of SHP. We enumerate all simulation- and implementation-critical parameters (e.g., pointer throttling, rehash budget, rounding window, ECC variants and exact overhead), derived from the reference scripts.

  \item We analyze the throughput–stealth trade-off both analytically (expected attempts and amplification) and empirically across LAN/WAN scenarios.

  \item We discuss defenses and the evolving arms race, including ML detection strategies that inform currently available countermeasures.

\end{itemize}

\section{The Silent History Protocol (SHP) Method}

\subsection{Threat Model and Basic Principles}

\subsubsection{Adversary Capabilities}
\ac{SHP} operates within the traditional threat model for a covert network channel, where an adversary monitors network traffic but does not have access to endpoints~\cite{Vivek2014Improving,Liu2013RealTime}. The adversary is assumed to have extensive network monitoring capabilities and actively searches for \ac{CC} by analyzing the characteristics of network traffic~\cite{GAS22,Ayub2019A}. In particular, the adversary aims to detect covert communication based on timing patterns by employing statistical detection thresholds, such as identifying timing deviations or unusual traffic volumes from certain systems \cite{Ondov2022Covert,Iglesias2017Are}.

Our evaluation and detectability analyses target this \emph{network-path based warden}. We discuss host-based defenses (e.g., endpoint log analysis or local heuristics) only as a separate defender capability in Sect.~\ref{sec:evaluation}; they are not part of the attacker model and do not influence the network-only assumptions used for analysis and experiments. We adopt this single attacker model as our coherent guideline throughout design, analysis, and experiments.

Traditional detection approaches for timing-based covert channels focus on identifying statistical anomalies in packet inter-arrival times or increased transmission delays caused by embedding hidden information \cite{Iglesias2017Are,Chen2015A,Djidjev2011Graph}, often combined with machine learning techniques~\cite{Darwish2019Using,Epishkina2019Timing}. Prior research suggests that a basic timing-based \ac{CC} can be detected if its manipulations result in an increase of the average delay by a factor of two or more~\cite{aleidi20}.
Recent advances have produced more sophisticated covert channels that evade detection. For example, the \textit{Epsilon-Kappa-libur}\cite{epskappa24} introduces an adaptive encoding mechanism that circumvents popular heuristics and degrades the performance of recent ML-based detection methods such as GAS and SnapCatch\cite{GAS22,snapcatch20}. In response, detection techniques have also evolved, with recent work using isolated binary trees advancing the detection of such sophisticated covert channels \cite{LIN2025104200}.

\subsubsection{Defense Strategy}
With \ac{SHP}, we aim to increase the amount of covert data that can be transmitted in a network for a given time period while remaining below detection thresholds. This requires that any covert signals appear as legitimate network behavior, indistinguishable from normal traffic timing patterns. 

To achieve this, we leverage the principle of channel amplification first developed in \ac{DYST}~\cite{DYST22}. Amplification minimizes the footprint of covert signaling by encoding information via small pointers to legitimate network events rather than modifying the timing or content of packets directly. This design philosophy ensures that even sophisticated statistical analyses would struggle to differentiate \ac{SHP} signals from normal network variations, as the actual modifications to traffic patterns are minimal compared to the volume of information transferred. Furthermore, by carefully calibrating the amplification factor, we can maintain a favorable trade-off between covert bandwidth and detectability risk.

Within this context, we assume that the communicating parties have a shared pre-established secret (such as keys or configuration parameters) that allows them to agree on how to interpret the timing patterns they observe. Furthermore, we assume that the overt traffic upon which the covert channel is built continues to flow naturally in the network and serves legitimate purposes for users unaware of the covert communication taking place. This approach distinguishes \ac{SHP} from more active covert channels that solely rely on generating their own traffic, potentially increasing detectability through anomalous network behavior.

\subsubsection{Operational Scenarios}
We consider two scenarios that typify the application areas of a covert network channel~\cite{Schneider2023Network,Heinz2020Covert}: LAN and WAN.

In the WAN scenario, we assume two nodes along a network path want to communicate covertly and can both observe all overt traffic passing through them, for example, because these nodes are two consecutive routers on a popular routing path. This scenario presents unique challenges related to network latency and jitter, which can desynchronize the timing observations between the communicating nodes. Additionally, WAN environments typically involve multiple autonomous systems with varied traffic patterns, allowing \ac{SHP} to blend its covert signals with diverse background traffic characteristics that might confound detection methods.

In the LAN scenario, covert agents filter overt traffic to use only packets that are visible to all network nodes as \ac{POI}, ensuring they can be observed by both the \ac{CS} and \ac{CR} when residing in the same broadcast domain network~\cite{Schmid23,dua2021covert}. The proximity of communicating nodes in this scenario offers lower latency and more predictable network conditions, but also poses different challenges as broadcast traffic volumes may be lower and more scrutinized by local security monitoring. Furthermore, local network segments often implement more stringent traffic filtering and analysis, requiring \ac{SHP} to adapt its parameters to maintain stealthiness while achieving acceptable bandwidth.

\subsection{SHP Algorithm}
\label{sec:SHP_algorithm}

\subsubsection{Basic algorithm}

\ac{SHP} uses timing information that is observable by all covertly communicating nodes as the input source for a history protocol. To improve the probability of matching a random bit distribution, the input value is deskewed before matching, e.g. by hashing it~\cite{cachin98}.

To achieve this, a message of length $m \in \mathbb{N}$ is divided into fragments of length $n$, $n \in \mathbf{N}, n<m$. Then the following steps are continuously performed until the message has been sent:
\begin{enumerate}
    \item Record legitimate traffic.
    \item Note the timing information of every \ac{POI} observable by both \ac{CS} and \ac{CR}. 
    \item Calculate data bits: Use the observed timing information to generate a bit sequence representing the input data, for example, by hashing the timestamp of the last received frame and taking the first $n$ bits of the hash.
    \item If the input bit sequence matches the next $n$ bits of the message to be transmitted, send a pointer signal.
\end{enumerate}

Receiver algorithm:
\begin{enumerate}
    \item Record legitimate traffic.
    \item Note the timing information of all \ac{POI}.
    \item Upon receiving a pointer signal, calculate the data bits, for example, by hashing the timestamp of the frame received shortly before the signal and taking the first $n$ bits of the hash as the next bits of the received message.
\end{enumerate}

For clarity, the operation of \ac{SHP} can be understood as a dual-channel architecture: The signal channel transmits synchronization and pointer information via subtle modifications to the headers of the standard protocol (for example, within \ac{ARP} requests). The data channel is the overt traffic that is observable to both covert agents. Note also that a covert message can be read by multiple \ac{CR}s, especially in a broadcast domain scenario, enabling multicast or broadcast \ac{CC}.

This process defines a probabilistic mechanism by which the sender encodes message fragments into observable timing patterns. Since each matching event depends on the stochastic occurrence of a specific bit sequence within legitimate traffic, we can now analyze the transmission rate by considering the likelihood and expected frequency of such matches. For a message fragment of length $n$, where each bit can be either 0 or 1, the total number of possible input sequences is $2^n$. Assuming that each of them is equally likely, the probability \( p \) of randomly matching a specific sequence in one attempt is \( \frac{1}{2^{n}} \). 

The expected number of attempts to hit the chosen sequence follows a geometric distribution~\cite{beria2006confidence}. The expected value of a geometric distribution with success probability \( p \) is \( 1/p \). Therefore, the expected number of attempts \( EA \) to hit the chosen sequence is:

\begin{equation}
\label{eq:expected_attempts_per_sequence}
EA(\text{n}) = \frac{1}{\frac{1}{2^{n}}} = 2^{n}
\end{equation}

So, on average, it will take \( 2^{n} \) observed packets to match the chosen sequence. A larger $n$ means that matches are fewer, but the payoff will also be more bits transmitted. The expected (average) number of bits transmitted per matching attempt is given by multiplying the payoff by the probability of that payoff:

\begin{equation}
\label{eq:bits_per_attempt}
\mathbb{E}_{bits}(\text{n}) = n \cdot \frac{1}{2^n} = \frac{n}{2^n}
\end{equation}

\subsubsection{Packets of Interest (POI)}
In the LAN scenario, only overt traffic that can reasonably be expected to be visible to all network nodes is used as \ac{POI}. For this, \ac{DYST} relied on Ethernet and \ac{IP} broadcast traffic, as well as IPv6 link-local multicast (e.g., \texttt{ff02::/16})~\cite{DYST22}. 

In \ac{SHP}, we expanded these input sources to incorporate a broader range of broadcast and multicast behaviors, such as STP and LLDP frames in layer 2, DHCP and IGMP on layer 3, and \ac{IP}v6 neighbor discovery and router advertisement messages. This expansion not only increases the volume of usable traffic, but also enhances the ability to operate under the radar in diverse network conditions.

In total, the following types of traffic are used:

\begin{itemize}
  \item \textbf{Ethernet broadcast and multicast}  
        Matches every frame whose destination MAC has the \emph{Individual/Group} bit set, i.e., the global broadcast
        \texttt{ff:ff:ff:ff:ff:ff} and \emph{all} multicast group addresses

  \item \textbf{IPv4 limited \& directed broadcasts}  
        Includes the limited broadcast \texttt{255.255.255.255} (DHCP, PXE, …) and every subnet-directed broadcast
        (last host address of the attached IPv4 network).  

  \item \textbf{IPv4 multicast}  
        Captures the entire 224.0.0.0/4 space (IGMP, mDNS, streaming, etc.).  

  \item \textbf{IPv6 multicast}  
        Matches everything in ff00::/8—router/neighbor discovery, MLD, mDNS, service discovery, …  

  \item \textbf{ARP (IPv4) \& NDP (IPv6)}  
        Neighbour-discovery protocols (with NDP already covered by \texttt{ip6 multicast})\\
\end{itemize}

However, integrating these extra protocols also entails certain drawbacks. First, the complexity of the code increases, making both configuration and maintenance more prone to errors. Secondly, a wider range of protocols is more susceptible to packet drops, which can potentially result in communication mismatches.

\subsubsection{Network-timing input sources}
The next step after filtering traffic of interest is to determine the bits of data to which we want to point. Using the right overt network pattern referred to by our pointers, designated as the input source, is vital for our timing \ac{CC}. The source could be any observable network-timing traffic characteristic like \ac{IPD}s, which can be somewhat consistently monitored by both the sender and the receiver.

In contrast to \ac{CC} based on actively manipulating time pattern, our algorithm will have to rely on a source of input bits that is already available on the network. It follows that, in order to optimize the data carrying capacity of the channel, it is essential to use a readily available input source with substantial informational density. In other words, the input source should have a preexisting high entropy as established by Shannon~\cite{shannon48}, since it quantifies the uncertainty or the average informativeness of a variable's possible outcomes. Therefore, entropy, given by

\begin{equation}
\label{eq:entropy}
H(X) = -\sum_{i=1}^{n} p_i \log_2(p_i)
\end{equation}

where $p(x_i)$ represents the probability mass function of the variable, serves as a crucial metric in evaluating the suitability of input sources for our purposes. This input source should also be relatively stable between \ac{CS} and \ac{CR}, so network timing discrepancies may need to be handled by rounding the timing information. 

\begin{figure}
  \centering
  \includegraphics[width=\linewidth]{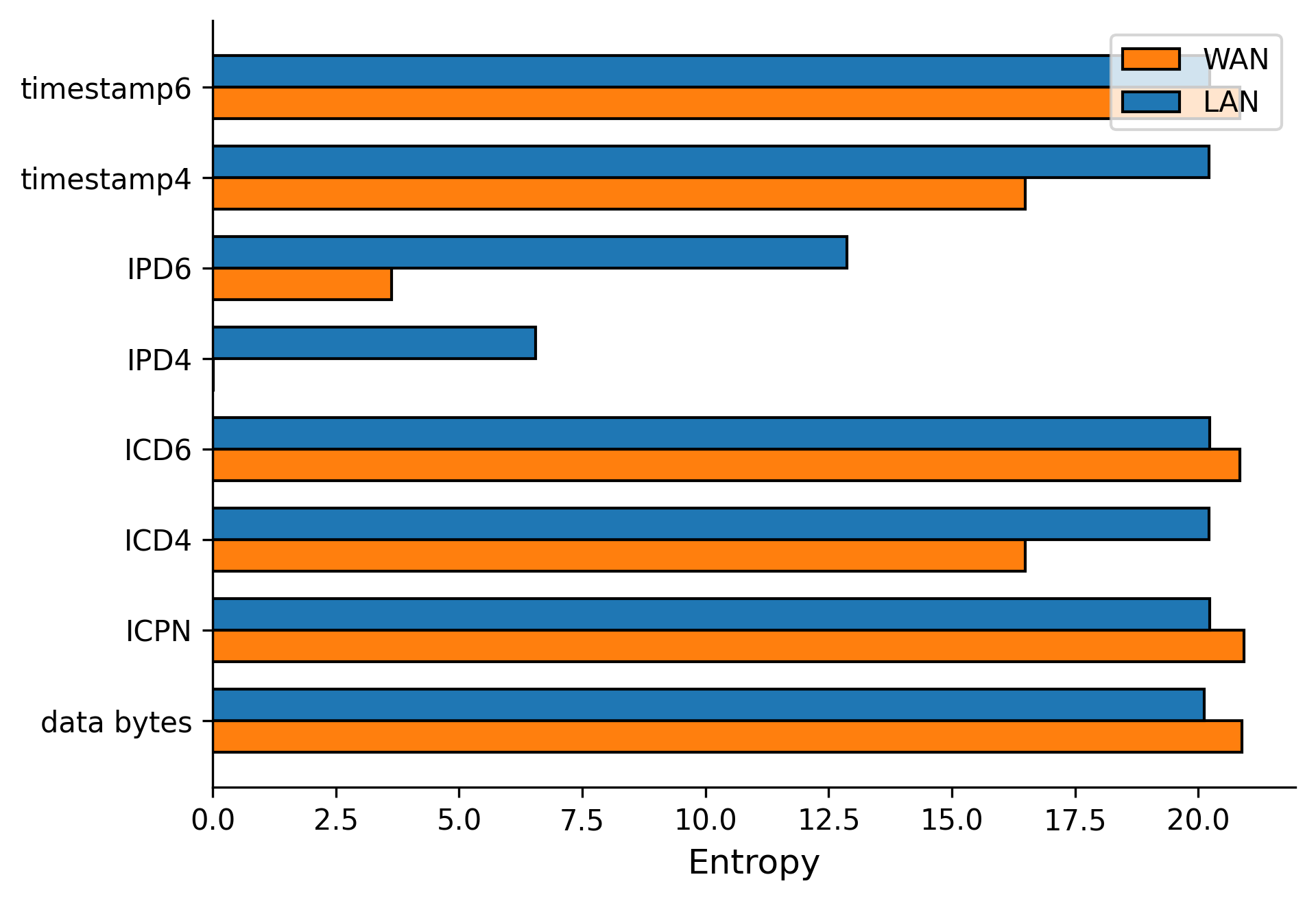}
  \caption{Measured entropy of different network-timing input sources. The numbers behind \ac{IPD}s and \ac{ICD}s indicate the applied rounding factor ($\epsilon$), where 6 represents no rounding and 4 corresponds to rounding to four decimal places. For comparison, the entropy of full packet data bytes—representing non-timing-based input sources—is shown at the bottom.}
  \label{fig:entropy}
\end{figure}

In order to calculate the entropy of various network-timing characteristics and find possible trade-offs when rounding, we utilized network traces from a home environment for the LAN scenario and the WIDE project for the WAN scenario. We tested for a wide range of possible timing input sources -- including \ac{IPD}s and system clock timestamp -- and rounded to whole seconds, tenths of a second, and thousandths of a second. Our analysis revealed that \ac{IPD}s exhibited low entropy, characterized by repetitive patterns close to zero, which makes it unsuitable for shuffled hash input values. In contrast, the reception timestamp measured by the system clock displayed high entropy, indicating its potential as an effective carrier for our \ac{CC} due to its greater variability. An overview of the results can be found in Figure~\ref{fig:entropy}.

Although clock timestamps have the necessary entropy, nearly identical timestamps between sender and receiver are an additional requirement for \ac{SHP} to function properly, because this assumes fine-grained clock synchronization between communicating nodes and limits communication to connections that can ensure minimal delays during transmission. This was also a limiting factor for \ac{DYST}, especially in the WAN scenario, where agents may not even observe \ac{POI} at roughly the same time, much less send and receive covert data within tight timing windows. Conceptually, embracing uncertainty in the observable timing, rather than fighting it via tight wall-clock synchronization, echoes recent covert designs that randomize transmission instants to maintain covertness under a warden’s binary test\cite{Lu2023TimeUncertainty}.

\begin{figure}[!htb]
  \centering
  \includegraphics[width=\linewidth]{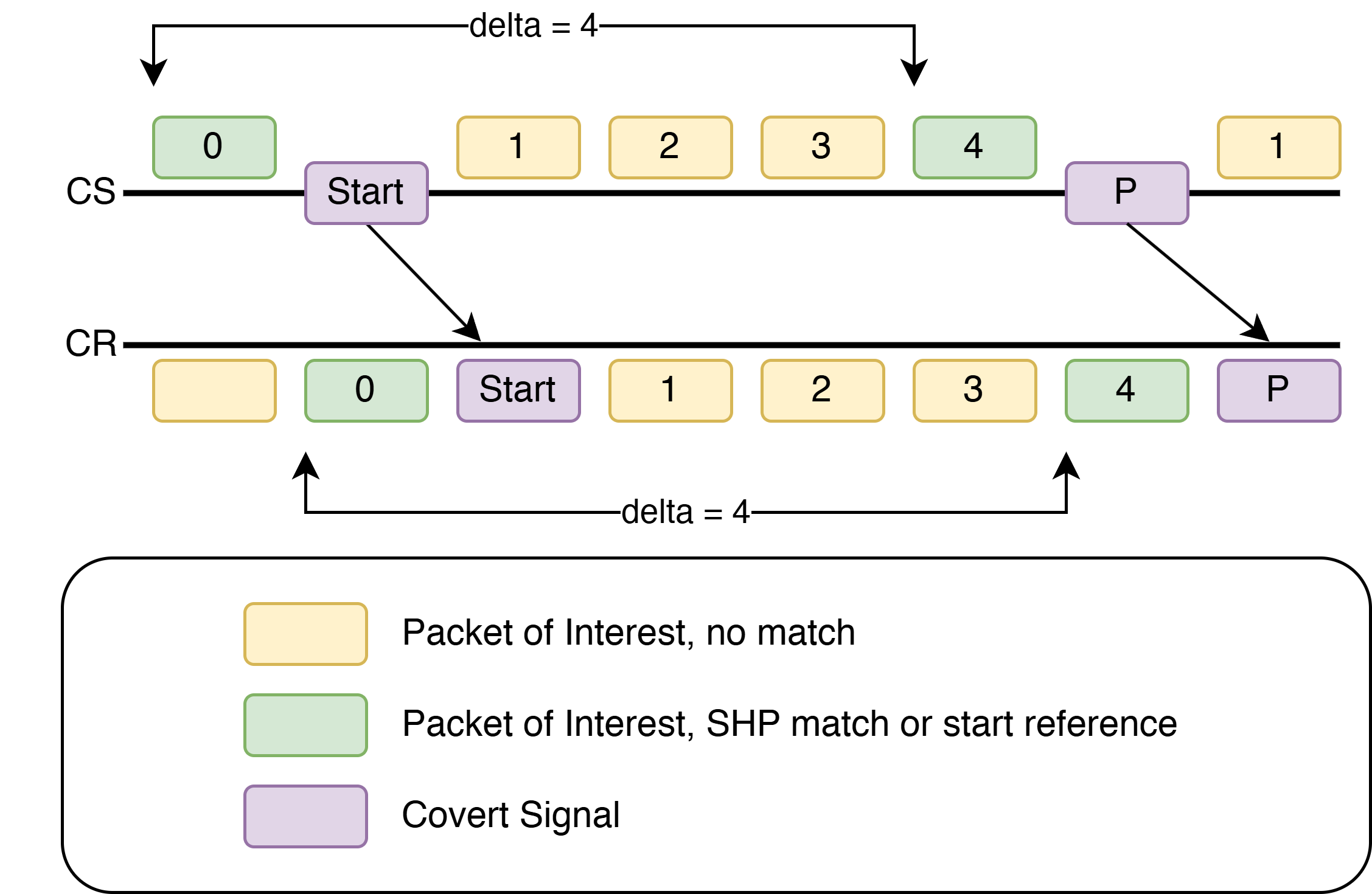}
  \caption{Covert transmission using relative input sources: The \ac{CR} measures the (time or count) delta between his reception of the start signal and the \ac{POI} marked by the pointer packet.}
  \label{fig:icd}
\end{figure}

To mitigate this issue while preserving entropy, we propose a novel approach: utilizing \emph{relative timing information} like the \ac{ICD}s, which is the time interval between connection initiation and the reception of the current \ac{PDU}. This involves sending an initial synchronization signal at startup and using its local clock timestamp as a reference point to calculate subsequent time intervals, as can be seen in Fig.~\ref{fig:icd}. By anchoring relative timing to this initial \ac{POI}, both the covert sender and receiver establish a common baseline that mitigates discrepancies caused by clock drift or variable network delays.

This approach simplifies synchronization by eliminating the need for continuous external time corrections, thus preserving the high entropy of the timing data while enhancing the overall robustness and applicability of \ac{SHP} across various network environments. This is especially true if timestamps need to be rounded, since this can easily nullify any \ac{IPD}s in congested environments. Because an initial synchronization signal is needed to initialize a connection regardless~\cite{cabuk06}, this approach is also not more detectable than using other sources of input data.

One caveat to using the relative timing approach in the LAN scenario is that more \ac{POI} might arrive at the \ac{CR} before the pointer has been calculated and transmitted. This would lead to the \ac{CR} interpreting the pointer as pointing to the wrong packet. Such a \ac{POI} mismatch is less problematic for \ac{SHP} than it was for \ac{DYST}, because we do not use the \ac{POI}s data bits for interpretation, but only their timing. As long as the measured timing differences between \ac{CS} and \ac{CR} stay within the allotted rounding window, the interpretation of the message will still be correct. To mitigate this issue further, several approaches can be used. First, more aggressive rounding leads to more tolerant matching. Second, instead of a naive implementation that uses the last \ac{POI} that arrived at the \ac{CR}, the pointer delay can be retroactively subtracted from the arrival time to guess the right \ac{POI}. This requires determining the round trip time between agents at the start of the connection and then subtracting $\frac{rtt}{2}$ at the \ac{CR} from pointer arrival times. Third, agents can ensure a sufficient \ac{IPD} between \ac{POI} by enforcing a silence interval, a method recommended by Wendzel et al.~\cite{DYST22}. Because \ac{POI} often occur in bursts, silence intervals lasting more than a few milliseconds substantially reduce the number of \ac{POI} used substantially. Lastly, we can actually remove the need to reference the right \ac{POI} entirely by directly using the \ac{IPD} between both pointers. Even when using this direct reference the matching on the \ac{CS} only happens when matching \ac{POI} are detected, so the timing characteristics of the covert channel stay anchored to overt traffic, ensuring undetectability stays stable. 

In networks where the exact timing of \ac{PDU}s is unreliable, but \ac{PDU}'s are usually received in the same order (e.g., satellite transmission), an alternative pattern called \ac{ICPN} can be used. \ac{ICPN} uses the number of \ac{PDU}s received as the input value. This removes the requirement for timing accuracy and enables multiple match attempts within one time window, but requires that none of the theoretically observable \ac{PDU}s are in fact lost between \ac{CS} and \ac{CR}. The entropy of \ac{ICPN}s and \ac{ICD}s were quite comparable to that of the system clock timestamps, even when rounding, as can be seen in Fig.~\ref{fig:entropy}. 

Instead of just using the covert start signal as a reference, the relative delta can also be measured against the \textit{latest} pointer during the connection. This also resynchronizes timing between the nodes, thereby reducing the impact of cumulative timing- or packet-ordering errors during connection. We call this method \ac{ISD}s for time-based input and \ac{ISPN} for packet order input. For \ac{ISD}s and \ac{ISPN}s, entropy cannot be determined a priori by using solely overt traffic packet captures, because the measurements here are interdependent on the live timing behavior of \ac{SHP} itself. However, we also implemented and evaluated these signal synchronization methods to determine their practicality.  

\subsubsection{Algorithm Extensions}
Apart from message, scenario, and subnet, the basic SHP algorithm is only parameterized by $inputsource$ and the number of bits per message, called $bitlength$. To enhance the robustness and performance of SHP, we refined the initial approach based on test results. These refinements focus on optimizing the \ac{CC}'s undetectability, data rate, and reliability under various network conditions. By incorporating adaptive error correction techniques and dynamic timing adjustments, the refined algorithm can more effectively maintain covert communication in noisy or congested network environments.

\paragraph{Error Correction Mechanisms}
\label{sec:ecc}
Modern wired networks generally exhibit low bit error rates; however, wireless networks face higher susceptibility due to signal attenuation, interference, and multipath propagation~\cite{jacobs2015}. Given the additional noise inherent in network timing, robust error correction mechanisms are essential for reliable transmission. 

Since an \ac{ECC} cannot only be used to detect but also correct errors, it can also be used to enhance the data rate by intentionally tolerating bit errors when matching data to pointers. We implemented a feature for \ac{ECC} using Hamming codes with an additional parity bit, which can correct 1-bit errors and detect 2-bit errors reliably. SHP in this variant will not only accept perfect matches between input source bits and message bits, but also matches that differ by at most 1 bit. The sender will also send a pointer signal whenever such a near-match is observed on the network. This decreases the distance between matches. However, it also means matching requires $bitlength+length(hamming)$ bits, which in turn lowers the probability of a match. Tolerant matching also increases the likelihood of message corruption, because 3-bit errors between \ac{CS} and \ac{CR} are easily interpreted as correctable 1-bit errors by the receiver. 

An alternative approach is to make the error correction independent of matching and instead treat error correction as part of the secret message itself. This prevents it from impacting the likelihood of matching, which decreases dramatically for every additional bit required per match. We implemented one such approach, called \emph{inline-hamming+}. Here, the covert message is divided into message fragments and Hamming error correction bits are added to each fragment. The error correction is then executed after a fragment has been received in total, which may require multiple pointers depending on the $bitlength$ used. 

The overhead in bits added depends on the number of message (= matching) bits used. For a fragment of \(n\) bits, the number of Hamming parity bits \(r\) is the smallest integer with \(2^{r} \ge n + r + 1\). Overhead fractions are: \texttt{hamming} \(= r/n\); \texttt{hamming+} \(= (r+1)/n\); \texttt{inline-hamming+}: no extra comparison bits per match, but the transmitted message is expanded by \(r+1\) after every \(n\) data bits. For the bitlengths tested this means: \(n=2\Rightarrow r=3\) (150\% / 200\% for hamming and hamming+ respectively), \(n=3\Rightarrow r=3\) (100\% / 133\%), \(n=4\Rightarrow r=3\) (75\% / 100\%), \(n=8\Rightarrow r=4\) (50\% / 62.5\%).

\paragraph{Out-of-order-delivery and Rehashing}
The \ac{SHP} algorithm sends signals to point to the traffic observed shortly before, which assumes a steady flow of overt traffic as its base. However, in networks scarcely used or where we filter only for broadcast \ac{POI}, overt traffic may not be sufficient to meet matching demand. We introduced two features to improve the data rate, \textit{\ac{OOOD}} and \textit{rehashing}.

\ac{OOOD} is based on sequence numbers, which are usually used to ensure that the data packets are reassembled in order after they have traversed the network. It sends a pointer signal whenever the input bits match the message bits of one of the next $2^{oood}$ message fragments. While this increases the likelihood of matching, it also requires sending at least $2^{oood}$ bits per pointer signal to communicate the sequence number to the \ac{CR}, which in turn reduces the \ac{CAF} or necessitates using a higher $bitlength$ for holding the \ac{CAF} stable. In scenarios where sequence numbers are necessary for reliable end-to-end transmission, this feature can increase the covert data rate without additional overhead when \ac{SHP} is used end-to-end.

Rehashing is a feature for \ac{HCC} first introduced in \ac{DYST}~\cite{DYST22}. Without this feature, a single hash is computed per \ac{POI}. Throughput can be enhanced by introducing multiple hashing attempts, allowing a secret message to be discovered by computing multiple hashes up to a given rehashing budget of bits $m$. This permits up to \(2^{m}\) of additional hashing attempts per \ac{POI}, icreasing the likehood of a match per \ac{POI}. This approach requires us to inform the receiver exactly how many hashing iterations are needed to find the data pointed to. Rehashing therefore also reduces the \ac{CAF} or requires increasing the $bitlength$ parameter to keep it stable.

\begin{figure}
  \centering
  \includegraphics[width=\linewidth]{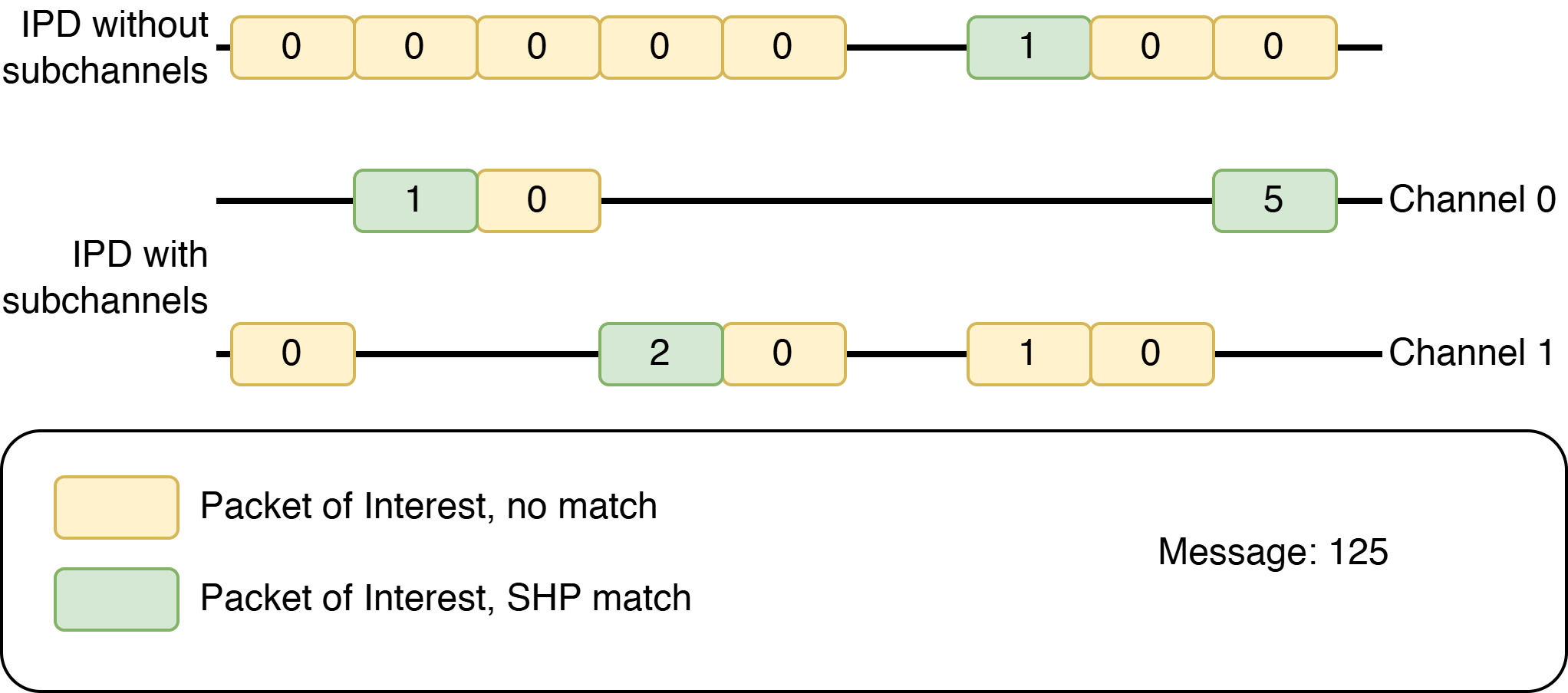}
  \caption{\ac{IPD}s without (top) and with (bottom) subchanneling. In congested networks, the \ac{IPD} is often close to zero. When traffic is divided into subchannels, input values become more diverse, which increases entropy.}
  \label{fig:explain_subchannel}
\end{figure}

\paragraph{Subchanneling}
As discussed, using simple \ac{IPD}s as input appears to have limited effectiveness, particularly in congested network environments, due to the frequent occurrence of intervals close to zero. To investigate whether this issue can be addressed, we enhanced matching by incorporating an additional layer of randomization observable by both \ac{CS} and \ac{CR}. This involves measuring \ac{IPD}s between multiple \ac{PDU}s instead of only consecutive ones. To achieve this in a deterministic manner, the \ac{CC} is divided into subchannels, as can be seen in Fig. \ref{fig:explain_subchannel}. Each \ac{PDU} is assigned to a subchannel based on a subset of its data bits, e.g., by hashing the source \ac{IP}. By measuring time deltas within each subchannel, we can achieve longer and more diverse intervals, thereby increasing entropy. Because the subchannel is determined for each \ac{PDU} in a deterministic manner, no additional bits are needed to signal the assigned subchannel to the \ac{CR}.

This approach is preferable to a naive implementation that would use the data bits as a source of additional entropy directly, because subchanneling is processed before rounding, so that \ac{IPD}s can be preserved as a somewhat useful differentiator in itself.

\subsection{Choosing parameters}
A \ac{CC} can be characterized by its position in the magic triangle of the three basic goals \textit{covert bandwidth}, \textit{robustness} and \textit{undetectability} \cite{fridrich1999applications}. The selection of optimal parameters depends on the specific constraints and goals of the covert communication scenario. If undetectability is paramount, the \ac{CAF} may be prioritized by increasing $bitlength$, which will in turn decrease the number of pointer signals required to transmit a given message. 

For scenarios where the signaling method allows for transmission of a predetermined number of bits of information per signal, it becomes crucial to determine the most efficient use of these bits to maximize bandwidth under the constraints of undetectability and error rates.

Because \ac{CS} and \ac{CR} do not always observe \ac{PDU}s at exactly the same time, simply using a timing value as exact as possible will lead to a vastly increased error rate. However, rounding the timing information aggressively will reduce the space of hash variability and therefore lead to fewer or even no matches. This suggests rounding the timing-based input sources to a value that can reasonably be expected to be identical between \ac{SHP} nodes.

\begin{figure}
  \centering
  \includegraphics[width=\linewidth]{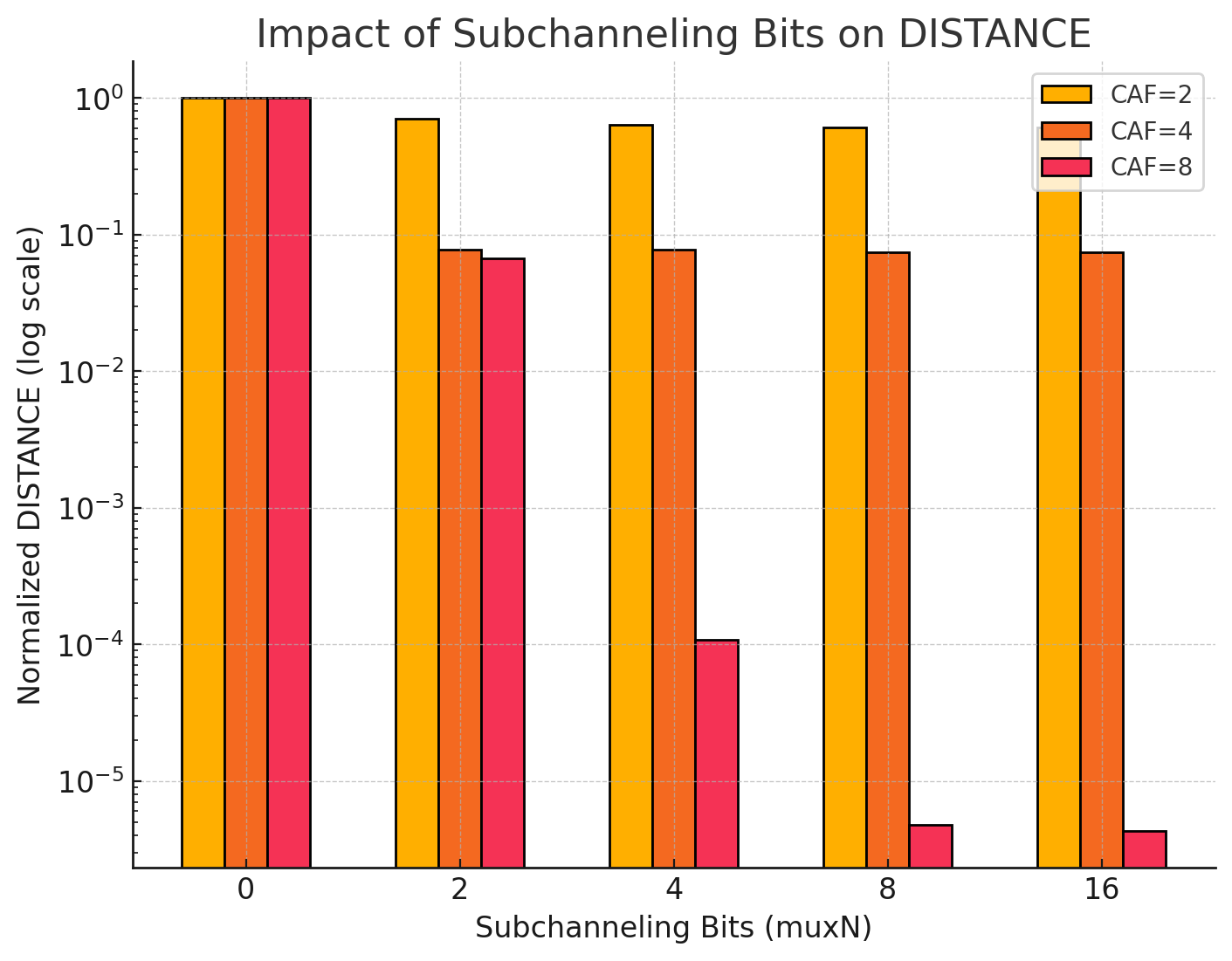}
  \caption{Relative distance between matches (x-axis) when using 0-16 bit for subchanneling (y-axis), for different amplification factors (normalized to no subchanneling, LAN scenario simulation)}
  \label{fig:subchanneling}
\end{figure}

The efficacy of subchanneling is illustrated in Fig.~\ref{fig:subchanneling}. It shows the average distance between matches for varying subchanneling bits. They have been normalized to the \ac{IPD}s distance measured  without subchanneling in the WAN scenario. The results indicate that deterministic subchanneling reduces the average signal distance while maintaining a consistent \ac{CAF}. The efficacy varies considerably depending on the entropy of the baseline \ac{IPD}s and other factors that influence the matching, such as the \textit{bitlength}. However, \ac{ICD}s still demonstrated superior matching performance in all tested cases.

The performance of \ac{ECC} variants is summarized in Tab.~\ref{tab:ecc}. We simulated transmissions with the message being the \textit{\textsc{universal declaration of human rights}} in a 90-second span and counted the number of transmitted message bits. Adding Hamming error correction bits to the matched message fragments decreased the likelihood of matching as it would if the bitlength were increased accordingly. This effect can be somewhat mitigated by adding a parity bit that allows for tolerant matching (hamming+). The best overall performance was enabled when error correction bits are treated as bits of the message (inline-hamming+), so that matching fragments are increased in number but not in the number of bits per match required. 

\begin{table}
  \centering
  \caption{Number of transmitted bits for a given message during 90s by ECC (WAN scenario offline simulation, ICD input source, all other parameters default)}
  \label{tab:ecc}
  \begin{tabular}{lrr}
    \toprule
    ECC type & CAF=2 & CAF=4 \\
    \midrule
    inline-hamming & 909 & 455 \\
    hamming+ & 397 & 253 \\
    hamming & 113 & 56 \\
    \bottomrule
  \end{tabular}
\end{table}

\begin{figure}
  \centering
  \includegraphics[width=\linewidth]{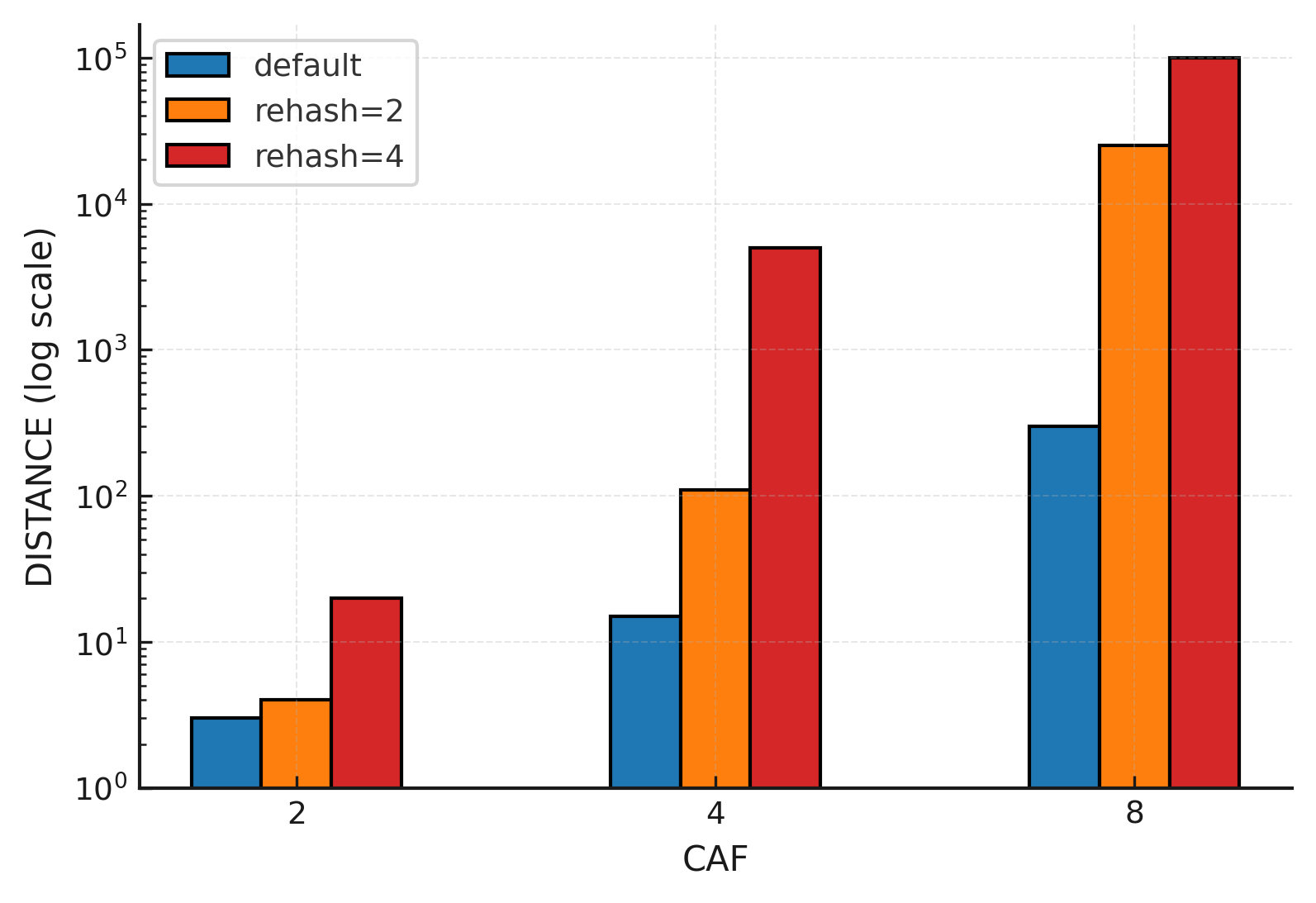}
  \caption{Distance between matches with and without repeated hashing for different amplification factors (LAN scenario simulation).}
  \label{fig:multipointer}
\end{figure}

\begin{figure}[h]
  \centering
  \includegraphics[width=\linewidth]{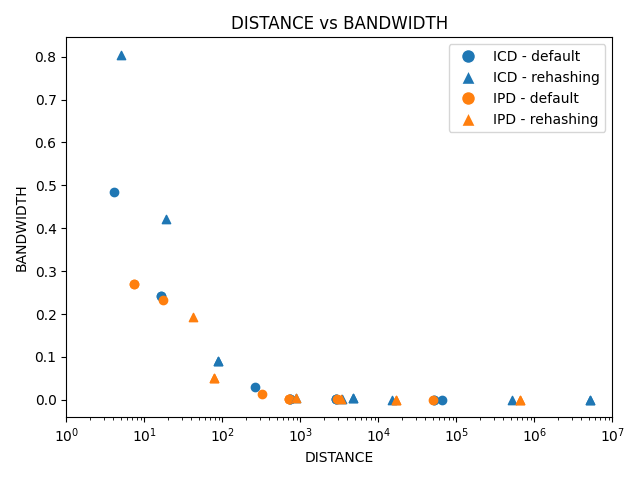}
  \caption{Distance and bandwidth for \ac{ICD}s and \ac{IPD}s with default settings (◯) and repeated hashing ($\triangle{}$), WAN scenario simulation, WIDE data from 2024-03-08, 2pm.}
  \label{fig:overall_performance}
\end{figure}

The trade-off between using multiple hash cycles and shorter bitlength is explored in Figs.~\ref{fig:multipointer} and \ref{fig:overall_performance}. When \ac{CAF} is held constant, using the default bitlength without repeated hashing results in a shorter distance compared to using multiple hashing retries, because holding the \ac{CAF} constant requires an increase in bitlength for matching. However, as can be seen in Fig.~\ref{fig:overall_performance}, employing repeated hashing increases the steganographic bandwidth for the same \ac{CAF}, as more bits are conveyed per match, even though matches are somewhat less frequent.

\section{Implementation}
\label{sec:implementation}
In order to test the feasibility of a covert channel via timing pointers, we developed two proof-of-concept Python scripts (server and client) that leverage software packet capture and real-time pointer injection. Python was used as a programming language because its high-level abstraction capabilities enabled us to express the complex parameterization of SHP concisely. The code is built around the Scapy framework to handle low-level network interactions and implements several tunable features such as silence intervals, \ac{ECC}, subchannel splitting, and multi-hashing. 

The scripts capture overt traffic and filter it for \ac{POI}. The only traffic manipulations conducted are limited to fractions of \ac{ARP} requests, which are used for covert signaling and pointers. Within ARP requests, we used the \ac{IP} range 127.55.0.0./16 for covert signals to make them easily discernible in the experiment. In a real environment, a less conspicuous address space -- one that conforms to the positions of agents in the network -- would be used. 
In the experiment, the \emph{third octet} was used to transmit rehashing bits. The \emph{fourth octet} contains a 6-bit watchdog checksum, as well as two bits that signal the type of message ($START$, $STOP$, or $DATA$). The watchdog checksum was used to experimentally verify the reliability of \ac{SHP} and its own \ac{ECC}, which in itself does not require these bits. 
Note that although we sent ARP requests, their content is legitimate, and we modify only a fraction of the IP address being requested so that the size of the pointer nested within the ARP request is smaller than the size of the secret message (for example, $CAF>1$). In a real environment, the requested ARP IP can be any legitimate address, as long as the \ac{CR} knows which address is to be used.
When mismatches are detected, a retry signal is sent over the covert channel so the sender will retransmit the last message fragment.

The scripts also enforce configurable silence intervals for both pointers and control signals to avoid collisions. A threaded design handles real-time processing, and the system logs key statistics (e.g., valid bits, ECC performance, average bit rate) for post-experiment analysis.

We also implemented an offline variant of \ac{SHP} that can use captured recordings of overt traffic to simulate parameter variations and various network conditions for our scenarios.

\section{Evaluation}
\label{sec:evaluation}
Using the implemented \ac{SHP} live and simulation scripts, we performed experiments to evaluate the performance and stealthiness of our proposed \ac{CC}. Covert bandwidth, robustness, and detectability were measured under various network conditions. In addition, we analyze the impact of the discussed parameters on communication performance.

Using the parameter space that can be found in Tab.~\ref{tab:config_params} we conducted experiments measuring \ac{CAF}, distance and steganographic bandwidth.

\begin{table}[t]
\centering
\caption{Configuration parameters and their value sets}
\label{tab:config_params}
\resizebox{\columnwidth}{!}{
  \begin{tabular}{@{}ll@{}}
    \toprule
    \textbf{Parameter}            & \textbf{Allowed Values}                     \\ \midrule
    \texttt{matching bitlength}            & \{2,3,4,8\}                          \\
    \texttt{rounding window $\epsilon$}      & \{1 s, 100 ms, 1 ms\}                                  \\
    \texttt{POI filter}                  & \{all, broadcast\_domain\}                  \\
    \texttt{POI reference}                  & \{direct\}                  \\
    \texttt{silence interval in ms}            & \{2\}                          \\
    \texttt{inputsource}          & \{IPD, ISD, ICD, ISPN, timestamp\}          \\
    \texttt{subchanneling\_mode}   & \{none, baseipd, iphash, clockhash\}         \\
    \texttt{subchanneling\_bits}   & \{0,2,4,8\}                                \\
    \texttt{ECC}                  & \{none, hamming, hamming+, inline-hamming+\}  \\
    \texttt{rehash bit budget $m$}            & \{0,2,4,7\}                                \\ \bottomrule
  \end{tabular}
}
\end{table}

The naive optimization domain of these variables contains 40.960 possible value combinations. After filtering out combinations that are contradictory (such as no subchanneling, but using subchanneling bits) and those with a $CAF \leq 1$, there remained 14,400 possible optimal combinations. 

Because analyzing large network captures is demanding in terms of computing resources and live experiments can only be conducted one after the other, we used genetic algorithms to find useful parameter combinations \cite{genai24}. The \ac{SGA} iteratively optimizes the selection of combinations of experimental parameters based on a performance metric computed from previous experiment runs called fitness. Each individual represents a distinct configuration of experimental parameters. At each iteration, the algorithm first selects a subset of elite individuals with the highest fitness scores. These elite configurations serve as parents in the crossover phase, where each offspring is generated by randomly inheriting parameter values from one of the two parents. To maintain diversity and explore new regions in the parameter space, a mutation operator is then applied to each offspring with a predefined mutation rate, potentially altering one or more parameter values. This straightforward evolutionary cycle—comprising selection, crossover, and mutation—enables the algorithm to progressively refine and schedule more promising parameter sets for subsequent experiments, while also ensuring that previously tested configurations are not redundantly re-evaluated.

The genetic fitness score was based on maximizing effective bandwidth, while requiring \(\text{CAF} > 1\), and then maximizing the product of the bit rate (\(\text{bps}\)) and the hitrate of \ac{ECC} matches for reliability:

\begin{equation}
\label{eqfitness}
\text{fitness} = 
\begin{cases}
0, 
& \text{if } \text{CAF} \le 1,\\[6pt]
\text{bps} \times \displaystyle \frac{N_{\text{ECC}}}{N_{\text{PR}}},
& \text{otherwise}.
\end{cases}
\end{equation}

Here, \(\textit{bps}\) is the bit rate, \(\textit{CAF}\) is the amplification factor (must exceed \(1\) to be considered valid),
\(N_{\text{ECC}}\) is the number of correct error-correcting-code matches and 
\(N_{\text{PR}}\) is the total number of pointer packets received.
This was complemented with manual sampling of likely candidates, such as low \textit{bitlength} and tolerant rounding to maximize bandwidth.

Live measurements were taken during the day and at night in a home network with 10 devices such as smartphones, smart home devices, and file servers. For offline analysis, traffic in this network was collected during the day and at night. In the WAN scenario, we used traffic recordings from the WIDE project~\cite{WIDE}.

\subsection{Environmental robustness}
\label{sec:robustness}

Robustness is crucial for covert channels to ensure reliable communication in real-world network environments where such impairments are common. This is especially true for timing-based channels, as network conditions affect packet timing and order~\cite{wendzelpatternsurvey15}. We evaluated the robustness of \ac{SHP} under various network conditions, specifically focusing on the impact of delay, jitter, noise, and packet loss on bandwidth.

\subsubsection{Impact of delay}

For \ac{SHP}, two kinds of delay need to be considered. \textit{Processing delay} is the time gap caused by the sender filtering packets of interest and determining to send a pointer. After that, \textit{network delay} refers to the time it takes for a pointer to travel from the sender to the receiver. If the application is highly optimized for processing (e.g., hardware-accelerated), networking delay typically becomes the dominant factor. 

In contrast, if processing involves complex deep packet inspection or inefficient code, processing delays may take precedence. The extent of this processing delay depends on the specific \ac{SHP} implementation, the system resources available to the \ac{CS}, and the bandwidth of traffic being parsed.

The observed values for one-way network delay are typically in the ranges $\leq 110$ ms in wired LANs~\cite{delay14} and $50$--$300$ ms in WAN communication~\cite{delay12}, depending on distance, routing, and network congestion. When using timing-based input sources, this causes differences in observed inputs between the \ac{CS} and \ac{CR}, especially when absolute timing is used as a reference for pointers. 

In the WAN scenario, consistent network delay has a reduced impact on the robustness of \ac{SHP}, as it can use relative timing (like in \ac{ISD}s) or packet ordering (like in \ac{ISPN}s) algorithms. This is because the \ac{CR} computes intra-signal delays relative to previous pointers or initial synchronization. Let $ISD_{CS}(n)$ be the difference between when the \ac{CS} observes a signal $n$ and the previous signal $n-1$, and the network delay between covert agents be $d$. Then the observed \ac{ISD}s for both parties are $ISD_{CS}(n)=(t_n+d)-(t_{n-1}+d)=t_n-t_{n-1}+(d-d)=t_n-t_{n-1}=ISD_{CR}(n)$. This means that the observed interpacket delays or relative times remain the same, rendering any consistent network delay inconsequential for matching in the WAN scenario. This aligns with recent covert communication work in continuous or time-uncertain settings, where robustness is also achieved by leveraging timing uncertainty rather than attempting to tightly synchronize clocks~\cite{Lu2023TimeUncertainty,Li2022Jammer}. Similarly, for the \ac{ISPN} algorithm, which relies on packet ordering, a constant delay does not alter the sequence of packet arrivals, so \ac{SHP} remains robust under high network delays as long as they are stable.

However, it is important to note that network delay can impact timing-based inputs in the LAN scenario because overt traffic is observed by both communicating parties at the same time. Furthermore, network delay also affects higher-level protocols whose performance depends on round-trip times, such as \ac{TCP}. If the delay is significant, it can also affect the overall throughput of legitimate traffic, which could indirectly influence the availability of suitable input sources for \ac{SHP}. Nonetheless, \ac{SHP} itself remains robust in various network lag scenarios.

To mitigate the effects of larger delays in the LAN scenario, a practical approach involves measuring the average delay between an overt packet and the pointer before initiating content communication. By communicating such a predetermined pointer delay adjustment to the \ac{CR} through a high-tolerance channel (i.e., pointing to a packet's data bits, which have less entropy, but are more stable), the \ac{CR} can adjust the timestamp of received pointers by subtracting the known pointer delay, ensuring better alignment between the pointer and the referenced packet. This preemptive calibration minimizes timing discrepancies, enhancing the reliability of the covert communication channel. The residual uncertainty of timing caused by differences in processing and transmission between \ac{SHP} operations can be considered jitter.

\subsubsection{Impact of jitter}

Unlike consistent delay, jitter can lead to unpredictable discrepancies in packet arrivals between the \ac{CS} and \ac{CR}, potentially causing mismatches in timing calculations and packet ordering. This can have a substantial impact on the performance of channels based on timing or packet order sources. For example in the context of \ac{ISD}s, jitter can cause timing mismatches, because the inter-packet delays observed by the \ac{CR} may differ from those observed by the \ac{CS}. 

\ac{HCC} can also experience pointer jitter, which refers to the variation in processing and network delay between the receiver observing an overt packet and its pointer. This leads to a decreased probability of identical input interpretation on both ends, thereby impacting the covert bandwidth. It can also change the order of overt packets between covert parties, so the \ac{CR} may interpret the pointer as pointing to the wrong packet.

To improve robustness against jitter, several methods can be considered: First, instead of requiring exact matches, \ac{SHP} can accept inputs that fall within a predetermined range of the expected value. These tolerance windows increase robustness at the cost of slightly reduced precision in the transmitted message. When using timing as input, this can be achieved using the timing rounding factor $\epsilon$. Increasing it improves robustness, as it increases the acceptable range of reception times, because the timing intervals are grouped into larger bins. For order-based input methods like \ac{ISPN}, we could interpret the $n$th packet as input $\lfloor \frac{n}{\epsilon} \rfloor$. However, tolerance windows also produce fewer unique input intervals, which decreases the entropy of the input source. This reduces matches and, with them, the covert bandwidth.

Second, frequent synchronization between sender and receiver can mitigate the effects of jitter. Techniques such as using start-of-frame indicators and silent intervals can help synchronize communication, ensuring that timing variations have minimal impact on data integrity. The \ac{ISD}s input method uses the last pointer as an indicator to synchronize timing. This neutralizes accumulating drift and prohibits the development of a consistent timing mismatch. 

We also implemented a silent interval using the method suggested by Wendzel et al.~\cite{DYST22}: If the \ac{CS} receives two or more \ac{POI} in less than $\phi$ milliseconds, all overt traffic in that time frame will be ignored, so that only isolated \ac{POI} will be considered for matching. Consequently, the \ac{CR} uses $\phi$ to ignore all \ac{POI} received less than this number of milliseconds before a pointer and only interpret it as pointing to \ac{POI} before that. This ensures the \ac{CS} can calculate and send a pointer without running into a race condition with consecutive \ac{POI}. Nonetheless, the larger we choose $\phi$, the more this will reduce matching chances and thus potential bandwidth. 

Third, the error correction codes we previously discussed need to be employed in order to catch remaining pointer mismatches, caused by jitter or otherwise. When discrepancies are detected, retransmissions or alternate pointers can be used to recover the lost information.

\begin{figure}[h]
  \centering
  \includegraphics[width=\linewidth]{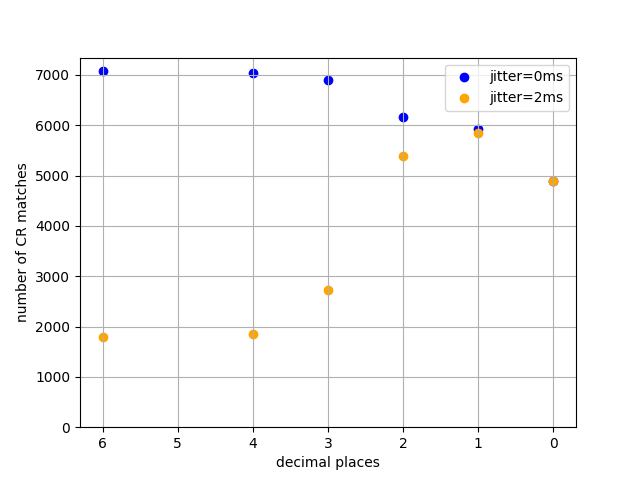}
  \caption{Impact of a 2ms jitter on the number of correctly received matches (y-axis) by rounding (x-axis), LAN scenario simulation, input source ICD.}
  \label{fig:jitter}
\end{figure}

We tested the impact of jitter using simulations as well as live network channels. In the simulations, we produced challenging network circumstances for the LAN and WAN scenarios, using a rather high time difference between \ac{CS} and \ac{CR} of 2ms for the LAN scenario~\cite{thombre20} and 20ms for the WAN scenario~\cite{jitter11} and only measured the impact of rounding network jitter of overt traffic. The results show that while less precise rounding improves performance when the jitter is negligible, more aggressive rounding improves performance when the jitter is higher. An example of this can be seen in Fig.~\ref{fig:jitter}, where for a 2ms jitter, rounding timestamps to one decimal place is optimal, while rounding has a negative impact on performance when jitter is close to zero. This also shows that, if optimal parameters are used, the impact of network jitter can be considered low in LAN scenarios, since jitter is typically below 1ms there.

The live network tests in the LAN scenario were intended to determine the interaction of all types of jitter and the other suggested measures. A naive test without any measures resulted in only 3 of 50 messages being transmitted properly, due to timing and \ac{POI} order differences between \ac{CS} and \ac{CR}, as well as the pointer arriving later than subsequent \ac{POI}. Using two short silence intervals and mild rounding as countermeasures yielded reliable transmission with the \ac{ECC} watchdog accepting $\geq 95\%$ of pointers, unless the network was congested and the scripts could not keep up parsing packets. If the \ac{ECC} of \ac{SHP} itself is used, reliability was $10\%$.
    
\subsubsection{Impact of packet loss}

Packet loss occurs when one or more packets fail to reach their destination, which can significantly impact \ac{SHP}, especially if the lost packet is a critical one in the form of a pointer or a matched packet.

\begin{figure}[h]
  \centering
  \includegraphics[width=\linewidth]{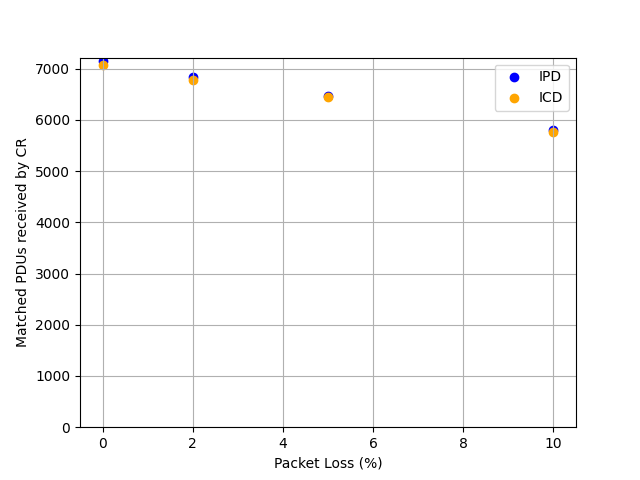}
  \caption{Impact of packet loss on the number of correctly received matches (y-axis) by percent of packets lost (x-axis), LAN scenario simulation.}
  \label{fig:packetloss}
\end{figure}

Packet loss presents a significant challenge to any communication protocol. Because in \ac{HCC}s data is only correctly received when both the pointer and the pointed-to packet arrive at the \ac{CR}, packet loss impacts protocols such as \ac{DYST} and \ac{SHP} more than overt traffic. In \ac{DYST}, packet loss could desynchronize the CS and CR, leading to errors in message reconstruction. As discussed,  \ac{SHP} incorporates options for error correction tailored for the covert channel to mitigate this risk. However, high packet loss rates can still decrease effective bandwidth of the channel, as can be seen in Fig.~\ref{fig:packetloss}. For example, a packet loss in the upper range of $10\%$, which can occur in tough WAN environments such as satellite communication~\cite{packetloss12}, reduces the effective bandwidth to around $72\%$. However, typical packet loss rates, which are below $2\%$, can be considered acceptable, as the steganographic bandwidth stays over $93\%$ under these conditions. 

SHP’s robustness could be further enhanced by integrating more sophisticated loss handling mechanisms that operate independently of timing, such as explicit resynchronization and acknowledgments. However, this would decrease round-trip-times. 

\subsubsection{Impact of overt traffic bandwidth}
\label{sec:overt_traffic_bandwidth}

The performance and reliability of the \ac{SHP} protocol are closely linked to the characteristics of overt network traffic. Since \ac{SHP} leverages existing network events to encode and transmit covert information, the amount and nature of overt traffic significantly influence the effectiveness of the channel.

The bandwidth was measured from two perspectives: First, \emph{the theoretically available} steganographic bandwidth ($sbw$), that is, the number of \emph{legitimate traffic bits} transferred divided by the average distance between matching PDUs. This value is informative for comparing the bandwidth of \ac{SHP}c with other \ac{CC} algorithms based on observed traffic like \ac{DYST} in general. Secondly, bandwidth was measured in traditional bits per second after error correction (= fitness). This measure is highly dependent on the specific network environment and traffic, but can be used to demonstrate the extent of \ac{SHP}s applicability in practice.

To quantify the impact of overt traffic bandwidth on \ac{SHP}'s performance, we conducted simulations using network traces with varying levels of traffic intensity as well as real-world tests. For the simulations, we utilized traffic captures from a home LAN environment during peak usage hours to represent high-bandwidth conditions and during night hours for low-bandwidth conditions. In addition, we simulate a WAN scenario using traffic traces from the WIDE project with different traffic volumes.

Furthermore, we performed live tests in our \ac{LAN} during peak and night hours under varying conditions, evaluating bitlengths from \emph{2} to \emph{64}. In each case, we examined both \ac{IPD}s and \ac{ISD}s as input sources, and we employed subchanneling in combination with rehashing techniques.

To test \ac{SHP}s performance under ideal conditions, we also tested a traffic condition of high packet rate. For this experiment, we generated live overt traffic designed to closely resemble high \ac{LAN} activity by emulating a mix of broadcast and multicast traffic types. Specifically, the live traffic generator produces ARP broadcast requests, IPv4 broadcast UDP packets, and IPv4 multicast UDP packets, each with randomized intervals between packets. The script employs a uniformly distributed jitter factor—ranging from 0.5× to 1.5× the mean interval—to avoid periodicity, with the default setting averaging around 120 packets per second. This results in a natural, non-deterministic traffic pattern, complete with variable source \ac{IP}s drawn from a specified subnet, that mimics the inherent variability and randomness observed in larger real-world LAN environments.

The live tests aimed to reveal how different parameter configurations impact matching, steganographic bandwidth, and reliability under diverse network loads. 

Our results indicate that \ac{SHP} performs optimally when the \ac{POI} filter and options like bitlength and rehashing are tailored to the amount of overt traffic observable by covert nodes. 

In low-traffic environments, we could achieve higher steganographic bandwidth using low bitlengths and rehashing, due to the increased frequency of matching opportunities. This holds if we require $CAF > 1$ to ensure \ac{SHP} has a positive impact on undetectability. The average distance between matches decreased below 800, leading to a higher data rate for the covert channel. This comes at the price of increased detectability. In LAN scenario live tests typical covert bandwidth came to around 0.1 bps when only \ac{POI}s visible to all nodes were used. 

On the other hand, in congested environments rapid matching leads to pointer collisions, and using all overt traffic makes pointing out the right packet harder. This potentially leads to mismatches and errors in the covert communication. Here filtering out overt traffic and using longer $bitlengths$ actually increased the effective bandwidth by leading to fewer mismatches.

Given the right amount of overt traffic and a quality connection, our Python scripts achieved a fitness of up to $265\;bps$ in live tests \textit{with} amplification, reached by \ac{ISD}s with $bitlength=8$ and $rehashing=7$. This far exceeds not only the performance of \ac{DYST}, but also Moss, another recent network history protocol optimized for throughput \cite{wang5634454moss}. Tab.~\ref{tab:high_low_bandwidth} summarizes the results using these parameters under different overt traffic conditions and compares them to \ac{DYST}\footnote{The \ac{DYST} life script was extended with the ability to record statistics and run with the parameters $chars=1,encoding=trivial$.}.

\begin{table*}[h] 
\centering
\caption{Impact of Overt Traffic Bandwidth on \ac{SHP} Performance (LAN scenario with 10 hosts, using ISD, $bitlength=8$ and $rehashing=7$, average of 10 runs)} \label{tab:high_low_bandwidth}
\begin{tabular}{lrr||r}
\toprule
Traffic Condition & Distance (SHP) & bps (SHP) & bps (DYST) \\
\midrule
LAN Peak Hours & 61.425 & 2.675 & 0.005 \\
LAN Night Hours & 66.037 & 2.383 & 0.004 \\
High Traffic (120 POI/s) & 10.549 & 121.276 & 0.887 \\ 
\bottomrule
\end{tabular}
\end{table*}

To validate the analytic model of \eqref{eq:expected_attempts_per_sequence} and \eqref{eq:bits_per_attempt}, we ran measurements on a LAN link (input source ISD) for bitlengths \(n\!=\!2,3,4\) both without sub-channelling and with four sub-channels.  For each setting we recorded the total number of observed packets until a match occurred and the number of bits actually delivered, then computed the realized bits-per-attempt and compared it to the theoretical value \(n/2^n\).  Across all six runs, the ratio  

\begin{equation}
\label{eq:phi}
\phi \;=\; \frac{\text{measured bits per attempt}}{n/2^n}
\end{equation}

ranged from approximately 0.70 to 0.99 (mean \(\approx0.83\)), indicating that our stochastic model correctly predicts covert throughput to within 20–30 \%.  The deviations can be attributed to residual timing correlations in legitimate frames and protocol-header processing overhead, but in all cases the experimental data remain well aligned with the exponential scaling predicted by \(EA(n)=2^n\) and \(\mathbb{E}_{bits}(n)=n/2^n\).

In summary, the overt traffic bandwidth has a significant impact on the performance and reliability of the \ac{SHP} protocol. High overt traffic volumes enhance the covert channel's bandwidth and robustness, while low volumes can hinder performance and increase detectability. If packet sequences are tight (low distance), filtering becomes necessary to avoid increased collisions, while in low traffic environments, \ac{SHP} can be configured to match more frequently. By understanding and adapting to the overt traffic conditions, covert agents can optimize \ac{SHP}'s effectiveness, maintaining a balance between covert bandwidth, robustness, and undetectability. Compared to \ac{DYST}, the performance gains of SHP are evident from our measurements under all conditions. Especially in high traffic scenarios, \ac{SHP}s adaptive configuration increase throughput in step.

\subsection{Detectability and Countermeasures}
Established methods for detecting network timing channels generally rely on comparing packet timings with expected traffic models~\cite{detect_ipd19,NIHBook,cabuk09}. However, these approaches appear unsuitable when applied to \ac{SHP}'s \emph{data} channel, because it does not alter packet data or their timings. Instead, defenders must rely on detecting the \emph{signal} channel that transmits the pointer.

Since our implementation of \ac{SHP} uses header bits within \ac{ARP} requests as the signal channel and since these bits could be modulated in arbitrary ways, the detectability evaluation will instead focus on detecting anomalies in the timing of these requests. We compare the distributions of interpacket gaps with and without \ac{SHP} channels. \ac{ARP}-specific detection methods, such as detecting replies without following traffic that uses the replied \ac{IP}-address, are excluded here, because other protocols could be easily substituted as \ac{SHP}-signaling channel (e.g., NTP~\cite{hielscher2021systematic} or HTTP~\cite{morepackets21}), but the \textit{timing} of these signals would need to stay quite constant if it is to be used as a pointer.

Applicable mathematical methods for detecting timing irregularities are statistical and information-theory tests. We used the Kolmogorov-Smirnov test for the former and the compressibility score introduced by Cabuk et al.\ for the latter~\cite{epskappa24,keller2023improving,cabuk06}. To evaluate the detectability of \ac{SHP}, we recorded real-world \ac{LAN} traffic with pointers using the parameters \textit{bitlength=8, inputsource=ISD} and used no extensions (out-of-order delivery, etc.).

\subsubsection{KS-test}
The Kolmogorov-Smirnov test is a statistical test for the similarity of two probability distributions~\cite{Iglesias2017Are,smirnov48}. It compares two empirical distribution functions \( F_{n_1}(x) \) and \( G_{n_2}(x) \) from two independent samples to determine if they come from the same distribution. The test statistic is the maximum absolute difference between these functions:

\begin{equation}
\label{eqkomolgorov}
D_{n_1, n_2} = \sup_{x} \left| F_{n_1}(x) - G_{n_2}(x) \right|.
\end{equation}

$D$ can then be used to determine if the difference between the two distributions is statistically significant by calculating the corresponding \textit{p-value}. Because this test does not assume a specific form for the underlying distributions, it can be used to compare them without assuming a particular traffic pattern~\cite{Grzesiak2023Covert,Zhang2020A}. 

We used the KS-test to determine if recordings of \ac{ARP} communication that include a \ac{SHP} channel can be differentiated from recordings without such a channel. Differentiation would be possible if the p-values between \ac{SHP}-recordings and non-\ac{SHP}-recordings are higher than the scores between non-\ac{SHP}-recordings. We applied the KS-test to pairs of traces taken from three corpora, each containing 10 traces for a message of $length \geq 1000~bits$:

\begin{enumerate}
\item \emph{non–non}: both traces contain only legitimate traffic;
\item \emph{\ac{SHP}–\ac{SHP}}: both traces embed an \ac{SHP} covert channel;
\item \emph{\ac{SHP}–non}: one trace carries \ac{SHP}, the other does not.
\end{enumerate}

All mutually distinct trace pairs inside each corpus were compared.  Tab.~\ref{tab:ks-results} summarizes the resulting mean KS statistic \(\overline D\), its standard deviation \(\sigma_D\), and the mean \(p\)-value \(\overline p\).

\begin{table}[ht]
\centering
\caption{Average two-sample KS results per corpus.}
\label{tab:ks-results}
\begin{tabular}{@{}lccc@{}}
\toprule
Comparison set    & \(\overline D\) & \(\sigma_D\) & \(\overline p\) \\ \midrule
non–non           & 0.0803 & 0.1076 & \(2.3\times10^{-5}\) \\
\ac{SHP}–\ac{SHP} & 0.0801 & 0.1074 & \(2.6\times10^{-5}\) \\
\ac{SHP}–non      & 0.0803 & 0.1074 & \(4.4\times10^{-5}\) \\ \bottomrule
\end{tabular}
\end{table}

The three mean KS statistics differ by less than \(2\times10^{-4}\), well inside one standard deviation. Likewise, all average \(p\)-values remain below \(5\times10^{-5}\), far under any customary significance threshold. Consequently, the test cannot distinguish traces that embed \ac{SHP} from those that do not. From the standpoint of a passive warden armed only with a two-sample KS-test, \ac{SHP} therefore remains indistinguishable from background traffic.

\begin{figure}[h]
  \centering
  \includegraphics[width=\linewidth]{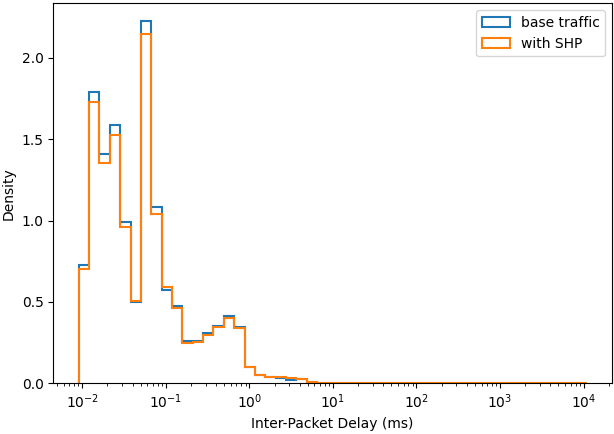}
  \caption{Distribution of inter-packet delays for two network captures (log scale), before and after filtering out \ac{SHP} communication, LAN scenario.}
  \label{fig:ipd_distribution}
\end{figure}

\begin{figure}[h]
  \centering
  \includegraphics[width=\linewidth]{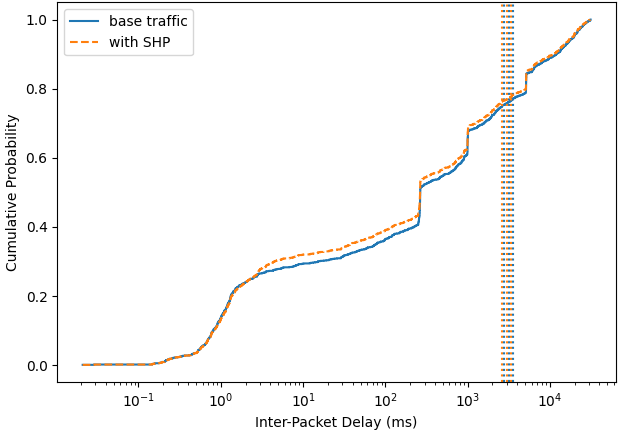}
  \caption{Cumulative distribution of inter-packet delays for two network captures (log scale), before and after filtering out \ac{SHP} communication, LAN scenario. Straight dotted lines represent the respective 95\% CI for average \ac{IPD}s values.}
  \label{fig:ipd_distribution_cumulative}
\end{figure}

\subsubsection{Compressibility Score}
The compressibility score is a statistical detection method that was first proposed for covert channels by Cabuk et al.~\cite{cabuk06,cabuk09}. This method evaluates the entropy of traffic parameters to identify patterns indicative of hidden communication. It operates on the principle that \ac{CC}s often introduce information structures into otherwise random traffic, thereby reducing its overall compressibility~\cite{epskappa24}. This approach is particularly useful for identifying anomalies in sets of information that inherently exhibit high entropy, such as \ac{ARP} timings.

The compressibility score for a given string \( S \) (formed by concatenating the \ac{IPD}s within a fixed-length window) is defined as

\begin{equation}
\label{eq}
\kappa(S) = 1 - \frac{|\Theta(S)|}{|S|},
\end{equation}

with $|x|$ representing the length of bitstring $x$ and $\Theta(S)$ being the compressed representation of $S$.

A higher value of $\kappa$ indicates that the data is more compressible, suggesting repetitive or predictable patterns typical of plain text or unencrypted data. Conversely, a value of $\kappa$ approaching zero implies that the data are less compressible, often characteristic of encrypted or random data with high entropy. 

To compute the compressibility score $\kappa$, we partitioned the recorded timing values into fixed-length windows, each containing 1,000 \ac{IPD}s. Within each window, the \ac{IPD}s were concatenated to form a string representation \( S \). We then applied a compression algorithm (\emph{GZip}) to compress \( S \).

\begin{figure}[h]
  \centering
  \includegraphics[width=\linewidth]{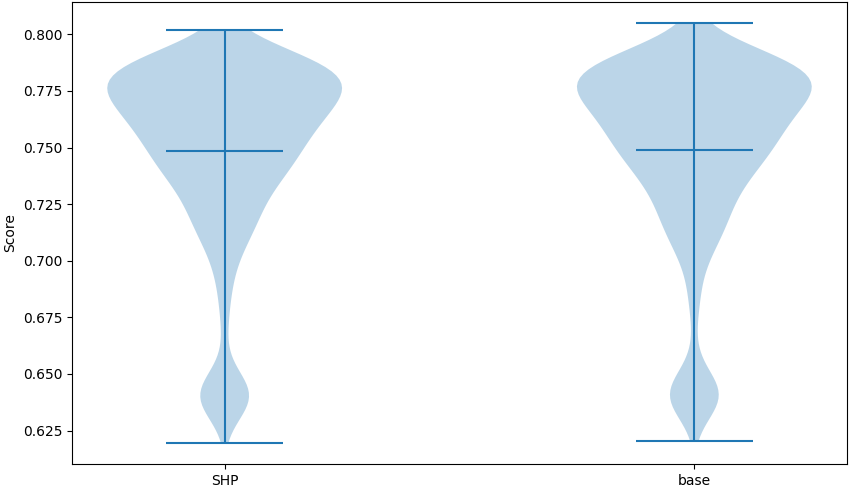}
  \caption{Violin plot of inter-packet delays for ten network captures over 1k bits, before and after filtering out \ac{SHP} communication, LAN scenario day and night.}
  \label{fig:plot_violin_compressibility}
\end{figure}

The violin plot presented in Fig.~\ref{fig:plot_violin_compressibility} illustrates the compressibility scores before (left) and after including covert channel packets (right). The distribution for both scenarios appears remarkably similar, with overlapping density profiles indicating minimal deviation between the two. The average $\overline{\kappa}$ of \ac{ARP} timing compressibility with \ac{SHP} pointers was $\overline{\kappa_{SHP}}=0.596$, while for traffic without a covert channel we calculated $\overline{\kappa_{base}}=0.594$. 

As can be expected~\cite{impactofequipment2012}, for different network captures, the overt timing distributions vary widely depending on factors such as time of day, network load or equipment, and protocol behavior, while the compressibility distributions between \ac{SHP} and the baseline \ac{IPD}s without covert channel stayed similar to each other. This suggests that \ac{SHP} packets can be introduced to the network without significantly affecting the compressibility of \ac{IPD}s even under varying conditions, implying that the covert channel remains largely undetectable through this metric. 

\subsubsection{Detectability by Request Frequency}
\begin{figure}
  \centering
  \includegraphics[width=\linewidth]{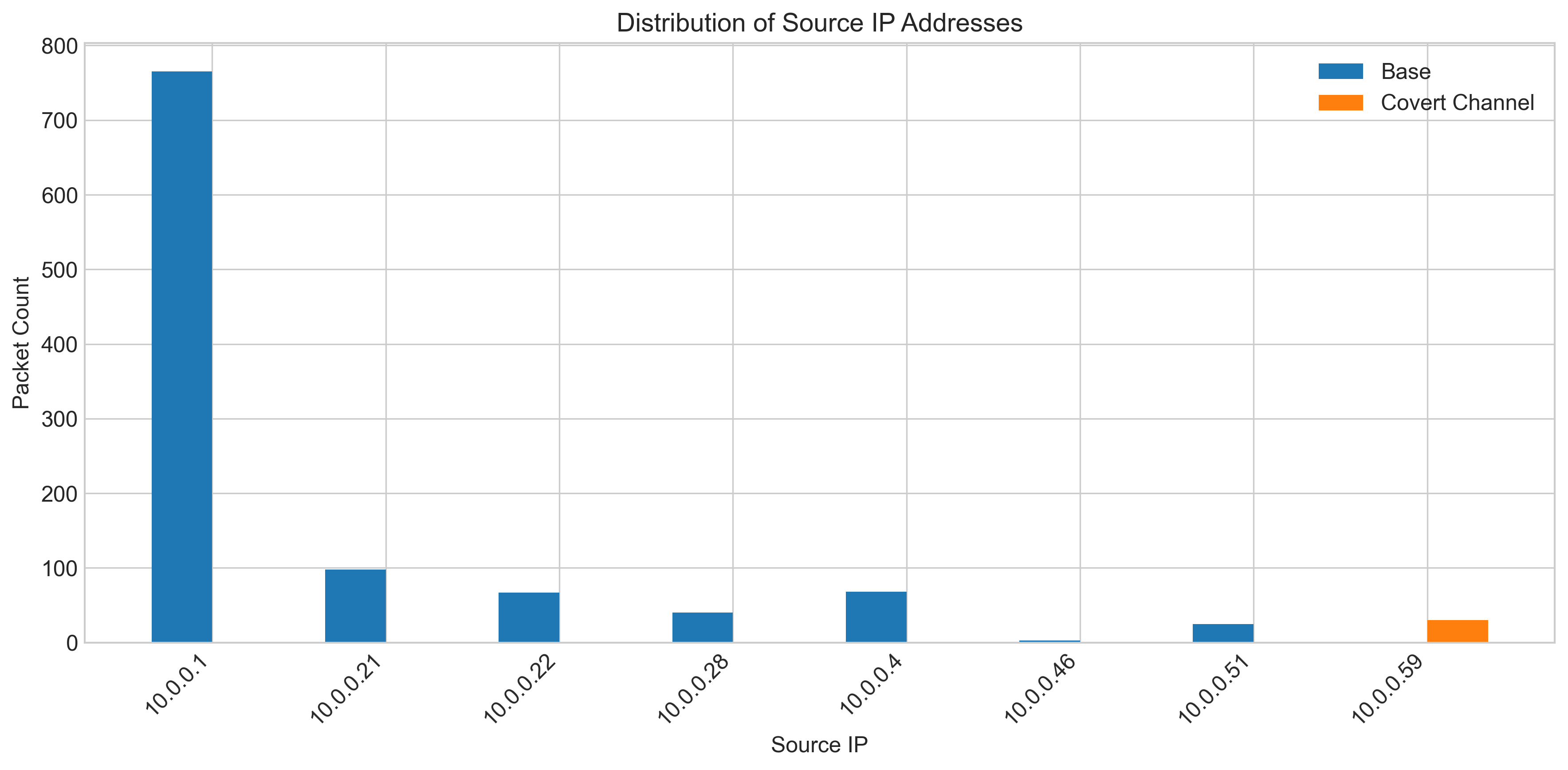}
  \caption{Frequency of ARP request by \ac{IP} source (LAN scenario, live capture). Frequencies of overt \ac{IP}s (blue) show high variability, so that the amount of covert pointers (orange) does not stand out. Since \ac{SHP} transmission is coupled with overt traffic its flow follows the networks.}
  \label{fig:histogramm_src_ip}
\end{figure}

A different approach to detecting a \ac{CC} using ARP requests is to consider the frequency of ARP requests either from a certain network node (mainly the \ac{CS}) or for a certain \ac{IP} or address space~\cite{Ondov2022Covert,Chen2015A}. We evaluated detectability using live \ac{SHP} capture in the LAN scenario using $input=ISD$ with $bitlength=8$. Because the probability of a pointer match occurring is \(1/2^8 = 1/256\), the additional request frequency introduced by the covert channel can be expected to be very low compared to the natural volume of legitimate network traffic. However, because \ac{SHP} does not only use \ac{ARP} as \ac{POI}, but uses \ac{ARP} as the signal channel, experimental verification is required.
Experimental evaluations using real network traces confirm that, under the given configuration, the proportion of covert pointer requests is negligible, even if only \ac{ARP} requests are considered by defenders. In several tests, the proportion of covert signals fell squarely within the typical range of legitimate traffic, making them effectively indistinguishable from normal background activity. An example can be seen in Fig.~\ref{fig:histogramm_src_ip}.

Furthermore, the stochastic nature of pointer generation under \ac{SHP} means that any additional requests occur in a random pattern rather than in a regular and easily discernible rhythm. Since channels mimicking the distribution of normal traffic are significantly harder to detect~\cite{Zhang2014Network}, the inherent conformity of \ac{SHP} to the randomness of the network traffic makes it even less likely for statistical anomaly detection methods to isolate it based solely on request frequency~\cite{noratelimiting11}. Because \ac{SHP} does not require using the requested ARP address for transmission (except for a few bits when rehashing is desired), requested addresses can be set to any desirable value to blend in with usual network activity. 

\subsubsection{Detectability by Channel Configuration}
We also examined the detectability of \ac{SHP} in high- and low-frequency signaling configuration to discern tradeoffs from this perspective. For this, we performed compressibility scoring using rehashing with \{0,4,16,64,256\} pointers allowed while keeping $CAF > 1$ and holding all other options stable.

\begin{figure}[h]
  \centering
  \includegraphics[width=\linewidth]{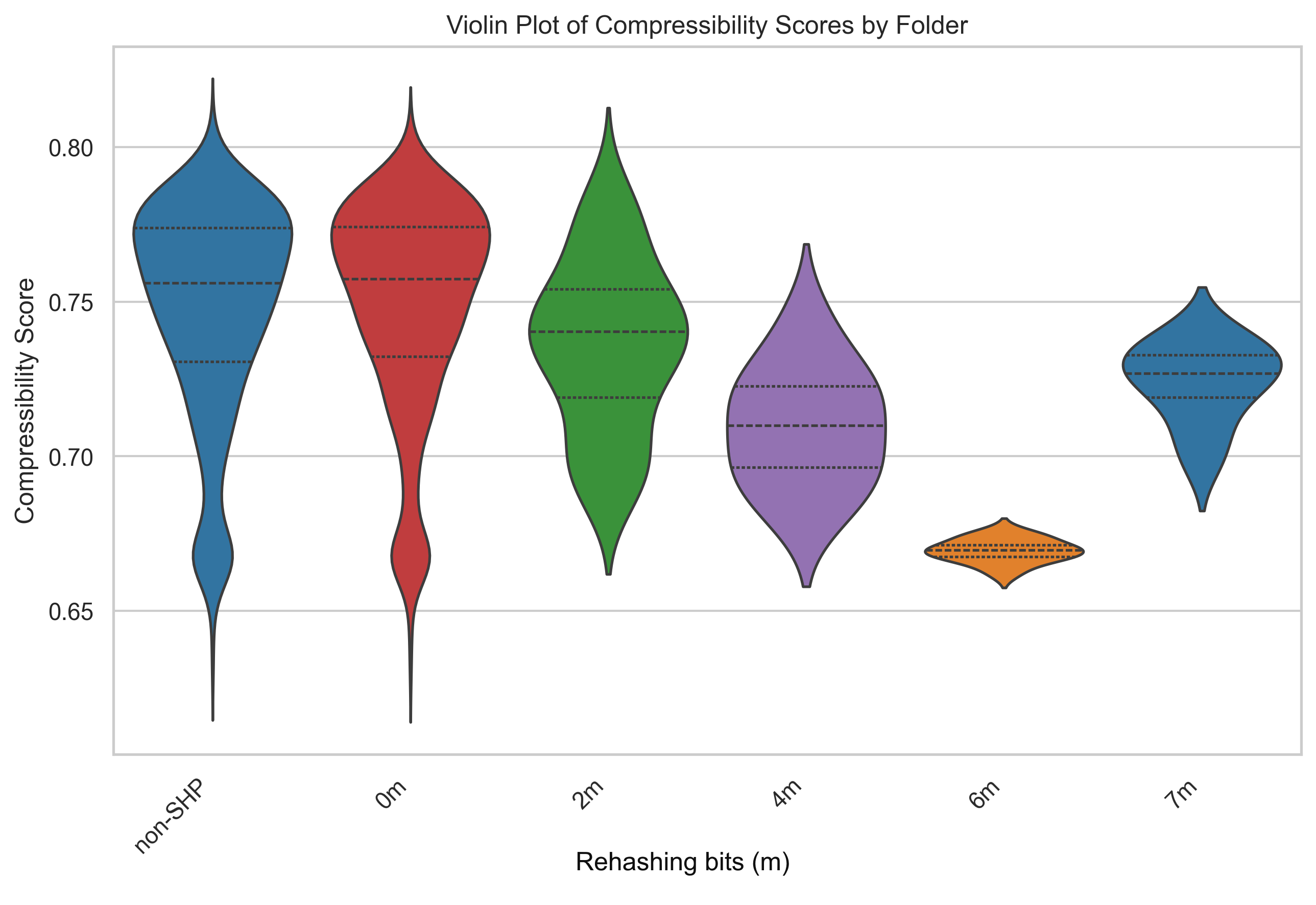}
  \caption{Violin plot of \ac{ARP} compressibility scores for different numbers of rehashing pointer bits (LAN scenario capture).}
  \label{fig:rehashing_violin}
\end{figure}

As Fig.~\ref{fig:rehashing_violin} illustrates, varying the rehashing budget (the maximum number $m$ of additional hash iterations that may be tried per overt packet) has a clear and systematic impact on the statistical footprint of the channel. Without rehashing ($0m$) the compressibility score distribution of the ARP‐request gaps is virtually indistinguishable from benign traffic (median $\tilde\kappa_{\text{non-SHP}}=0.756$ versus $\tilde\kappa_{0m}=0.757$, $p=0.55$). As soon as two or more extra hashes are permitted, however, the score begins to drift downward – indicating that the inter-arrival pattern becomes more regular and therefore more compressible. The medians fall to $0.740$ (2 m), $0.710$ (4 m), and $0.670$ (6 m) and all differences from the baseline are statistically significant at $p<0.01$. Interestingly, pushing the budget to seven bits (7 m) reversed part of this trend ($\tilde\kappa_{7m}=0.728$), a consequence of the highly random nature of timing distributions as well as the fact that highly frequent pointers begin to blend into the timing distribution of overall overt traffic. Traffic of such magnitude could be be detected by defenders using other means like request frequency detection. 

\begin{figure}[h]
  \centering
      \includegraphics[width=\linewidth]{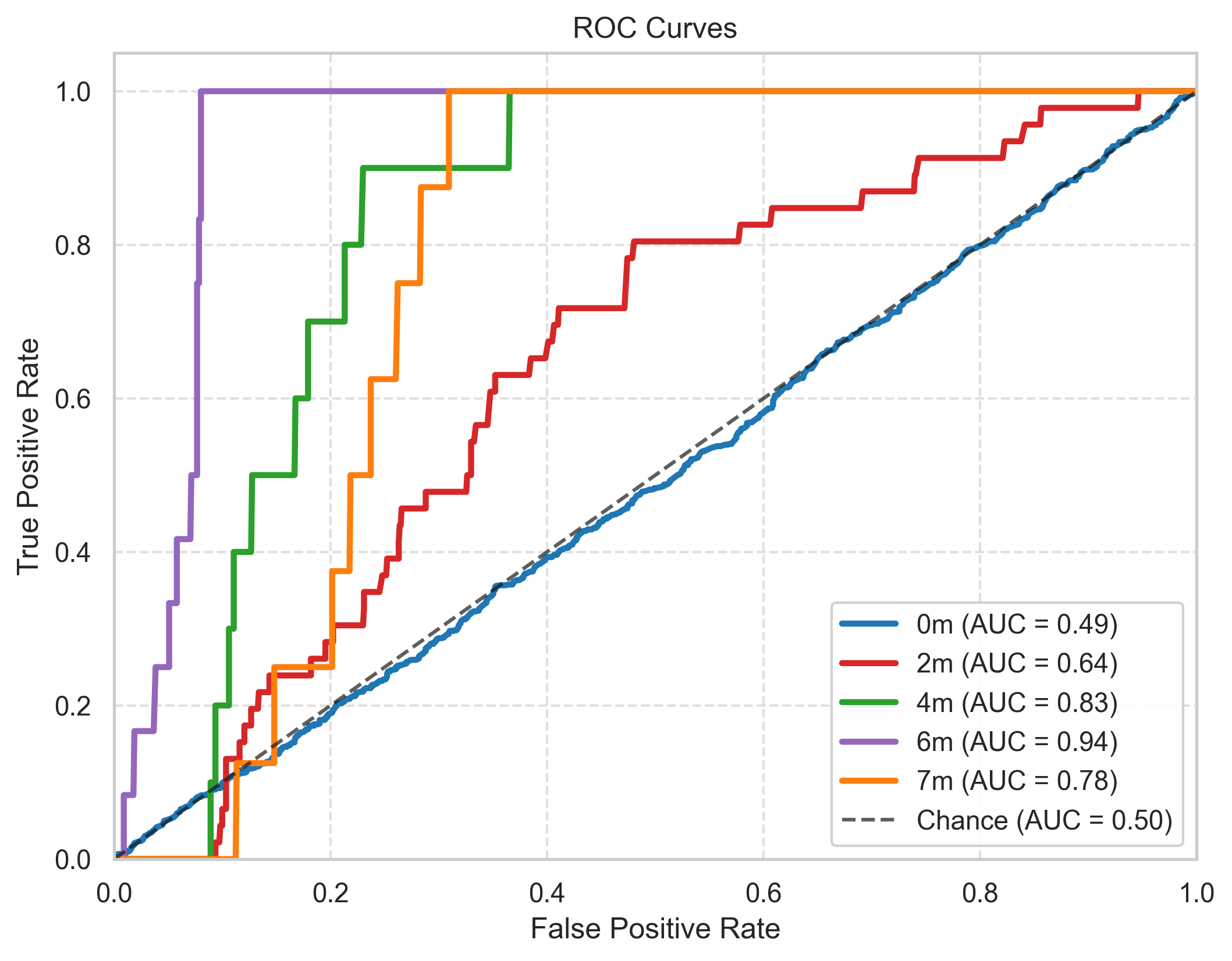}
  \caption{Detection performance based on compressibility scoring for different numbers of rehashing bits (LAN scenario capture).}
  \label{fig:rehashing_roc}
\end{figure}

The Receiver-Operating-Characteristic analysis in Fig.~\ref{fig:rehashing_roc} confirms this intuition from a defender’s perspective. A detector that used the windowed compressibility score performs no better than chance for $0m$ (AUC$\approx$0.49) and only marginally better for $2m$ (AUC$\approx$0.64). Once the budget reaches $4m$, however, the same test becomes much more effective (AUC$\approx$0.83).

Taken together, the results highlight an important configuration trade-off. Rehashing dramatically improves throughput (Sect.~\ref{sec:overt_traffic_bandwidth}) but also makes the channel susceptible to very lightweight statistical detectors. In the LAN traces studied here, keeping the rehashing budget below four extra hashes (i.e., $m<4$) preserves the compressibility profile within the natural variability of ARP traffic and forces a warden to resort to far more sophisticated, multi-feature analyses for reliable detection.

\subsubsection{Detectability with a GAS-style Model}
To assess detectability using machine learning, we tested whether \ac{SHP}’s \ac{ARP}-based timing channel could be identified by a detector based on GAS, a tensor-based method for detecting covert channels that exfiltrate data through small-volume transmissions \cite{GAS22}. The model was trained \emph{in-domain} on legitimate ARP \ac{IPD}s \cite{GAS22}. This isolates the signal channel and avoids confounding domain shifts that arise when models trained on flow \ac{IPD}s (e.g., HTTP/SSH) are applied to \ac{ARP}.

First, we transformed continuous \ac{IPD} sequences into a 16-state discrete alphabet using context-aware variation encoding. For each center position $i$ in the \ac{IPD} sequence, we computed forward and backward derivatives on aligned center points: $f_i = \mathrm{IPD}_{i+1} - \mathrm{IPD}_{i}$ and $b_i = \mathrm{IPD}_{i-1} - \mathrm{IPD}_{i}$. We then form a 4-bit code from the signs $\mathbb{1}[f_i > 0]$, $\mathbb{1}[b_i > 0]$ and magnitudes $\mathbb{1}[|f_i| > m_f]$, $\mathbb{1}[|b_i| > m_b]$, where $m_f$ and $m_b$ are the medians of $|f|$ and $|b|$ computed on the training set (legitimate \ac{ARP} only). These medians are reused for validation and test to prevent discretization leakage. Then we applied quantile clipping of extreme values $[0.5\%, 99.5\%]$ to reduce the impact of extreme outliers introduced by capture artifacts or rare bursts\footnote{Disabling clipping did not materially change our conclusions.}. 

As ML-model, we adopted GAS's sequence prediction architecture: an embedding layer over the 16 discrete states, followed by a single LSTM layer and a dense softmax output over the next-state vocabulary. We use a sequence length (time step) of 8, an embedding dimension of 100, and 64 LSTM units, consistent with \cite{GAS22} and effective for \ac{SHP}'s sparsity regime.

\begin{figure}
  \centering
  \includegraphics[width=0.95\linewidth]{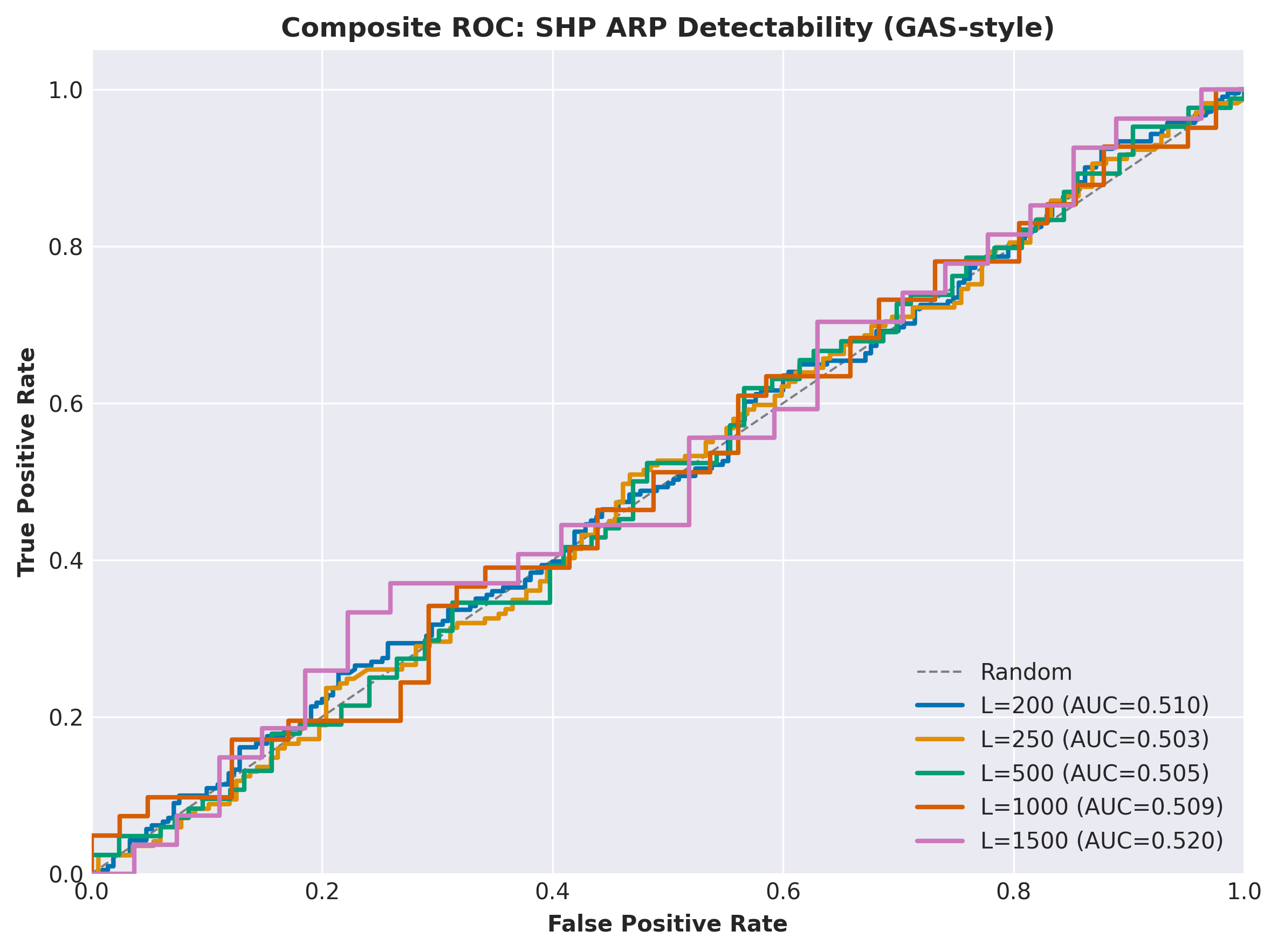}
  \caption{Composite ROC curves for $L\in\{200,250,500,1000,1500\}$ using the in-domain GAS-style model. Legends indicate AUC per curve.}
  \label{fig:gas-arp-roc-composite}
\end{figure}

As can be seen in Fig.~\ref{fig:gas-arp-roc-composite}, across $L \in \{200, 250, 500, 1000, 1500\}$, the ROC AUC remained close to chance with tight-to-moderate 95\% CIs and very low TPR at 1\% FPR (near zero).

\subsubsection{Countermeasures} 
As history protocols such as \ac{SHP} challenge available detection methods, they could alternatively be curbed by countermeasures that aim to disrupt the channel without detecting it. Such countermeasures fall into three primary categories: protocol normalization, rate-limiting and basic defense mechanisms.

\paragraph{Traffic Normalization and Timing Perturbation}

One of the primary defenses against timing-based covert channels is the implementation of traffic normalization techniques~\cite{normalization20}. Protocol normalization aims to standardize network traffic characteristics to eliminate timing-based patterns. By adding jitter or enforcing fixed inter-packet intervals, the entropy of the timing information is reduced, making it more difficult for the covert sender and receiver to find matching patterns.

However, this approach introduces latency and degrades the quality of service for legitimate applications, particularly those sensitive to timing, such as VoIP or real-time streaming~\cite{Epishkina2019Timing}. For example, certain real-time applications rely on timing characteristics for synchronization or performance tuning, which normalization may disrupt~\cite{2008On}. Because of its amplification and variety of input sources, \ac{SHP} can be configured to be robust while keeping the impact on bandwidth low as shown in Sect.~\ref{sec:robustness}. This means aggressive normalization would be necessary to disrupt communication. Moreover, \ac{SHP} also supports input sources that are not based on \ac{IPD}s such as packet order, so defenders would have to employ a wide range of normalization techniques to hamper transmission.

\paragraph{Rate Limiting}
Limiting the frequency of specific protocols, such as ARP or multicast queries, can disrupt \ac{HCC}s that rely on frequent pointer signals to legitimate traffic. However, protocols like ARP exhibit traffic patterns that are inherently unpredictable due to their reactive nature, making it difficult to establish meaningful thresholds for rate limiting even traditional \ac{CC}~\cite{Sun2021Suspicious,noratelimiting11}. Because \ac{SHP} enables amplification of covert communication, rate limiting its communication without considerable impact on overt network usage is even less feasible. 

\paragraph{Basic Defense Mechanisms}
Traditional cybersecurity defense measures can also be applied to limit the effectiveness of network channels like \ac{SHP}. Network segmentation can prevent broadcast or multicast communication between endpoints and minimize shared traffic accessible to \ac{HCC}s~\cite{Kogos2021Prospective,Ondov2022Covert}. Nevertheless, adaptive probing algorithms that analyze the communication methods available to them could pick from the parameter set of \ac{SHP}, e.g., input sources and POI, to enable peer-to-peer communication over those boundaries as long as any traffic at all is still allowed~\cite{Yarochkin2009Introducing,Bortolameotti2019VictimAware}.

Requiring host-based detection for all clients in a network enables detection not through the network characteristics that we have shown to be difficult to detect, but through local detection methods such as signatures or host-behaviour heuristics~\cite{Ondov2022Covert,Joe2021HostBased}. However, such measures are not any more effective against \ac{SHP} than they are against other network-based \ac{CC}.

\section{Discussion}
\label{sec:discussion}
The robustness of \ac{SHP} is a critical aspect that determines its viability in real-world network environments. Our evaluation shows that \ac{SHP} enables an elevated level of resilience against delay and can be configured to be robust under typical packet loss and jitter conditions. In \ac{DYST}, even slight variations in synchronization could lead to mismatches, drastically reducing the reliability of the channel. SHP addresses this by utilizing direct synchronization and relative timing, i.e., using \ac{ICD}s or \ac{ISPN}s to create a channel less dependent on its environment. Using relative timing within a connection, \ac{SHP} reduces the need for centralized synchronization.

Another critical factor influencing robustness is network congestion. Although increased traffic density can enhance the effectiveness of \ac{SHP} by providing more potential reference points, extreme congestion can introduce timing noise and mismatches. The inclusion of dynamic error correction techniques and resynchronization strategies (such as \ac{ISD}s recalibration) further strengthens \ac{SHP}’s ability to adapt to varying network conditions.

Despite these improvements, \ac{SHP} is still subject to environmental constraints, particularly jitter and the amount of available overt traffic. Up to a point, susceptibility to jitter could be mitigated by rounding the timing information and frequent synchronization. More complex algorithms may be necessary to further minimize this, for example, weighted timing inputs where the average or median of multiple observed timings within a time window is used for matching. 

The most fundamental limitation of \ac{SHP} is that its bandwidth and accuracy depend on the amount and timing of overt traffic which can be observed by communicating agents. This limitation can be alleviated by employing adaptive algorithms that dynamically adjust the \textit{bitlength}, {rehashing} and \textit{ECC} strategies based on observed network traffic conditions.  Implementing \ac{SHP} closer to the hardware, for example, using kernel access, could further improve performance by minimizing delays. For environments that do have a lot of overt traffic, a possible improvement is channel multiplexing: dividing overt traffic into multiple parallel channels increases spacing per channel, thereby deconflicting pointers. These multiple channels could transmit covert traffic in parallel or for different applications, leading to a substantial increase in overall bandwidth. Optimization of such a multi-channel method could essentially follow Wendzel and Keller \cite{wendzelLowAttention}.

While using a tailored \ac{ECC} can enhance bandwidth and still effectively mitigate some transmission errors, future research might explore the application of more advanced error-correcting algorithms, such as turbo codes or LDPC codes, which are well-suited for high noise environments. Moreover, the concept of multiplexing, although promising, remains underexploited. Developing dynamic channel selection and frequency-hopping techniques, similar to frequency-hopping spread spectrum (FHSS) in wireless communications, could further enhance the robustness and undetectability of covert communications by optimizing bandwidth usage and improving evasion capabilities in diverse network conditions.

Detectability evaluation has shown that a history \ac{CC}'s greatest weakness is its reliance on legitimate traffic and becomes a strength when it comes to covertness: The frequency of \ac{SHP} signaling rises and falls automatically with the frequency of legitimate traffic. This innate characteristic helps the protocol to stay under the detectability threshold given by the configured \ac{CAF}. However, continuing evasion also means being able to adapt to new detection mechanisms. As network traffic analysis becomes more advanced, detecting more subtle variations in timing or order may become possible. Future iterations of \ac{SHP} could improve their undetectability by incorporating obfuscation techniques, such as randomizing their pointer patterns. For now, detecting and mitigating \ac{HCC}s such as \ac{SHP} requires a costly multilayered defense strategy. More research on context-sensitive adaptive countermeasures will be critical to stay ahead of this growing challenge in cybersecurity.

To optimize \ac{SHP} performance across varying overt traffic conditions, the \ac{CS} and \ac{CR} could employ adaptive strategies. For instance, by continuously monitoring network jitter and dynamically modifying transmission intervals, these mechanisms could reduce the likelihood of errors. To this end, the \ac{CS} could potentially dynamically adjust the \textit{bitlength}, \textit{rounding factor}, and {rehashing} mechanisms based on real-time observations of the overt traffic bandwidth. In some cases, machine learning techniques might be utilized to predict traffic patterns and adjust parameters proactively. Additionally, in extremely low-bandwidth environments, the \ac{CS} may consider generating own legitimate looking overt traffic to create more matching opportunities, although this approach increases the risk of detection and would not be a pure history covert channel anymore.

\paragraph*{Ethical Statement}
While \ac{SHP} (like all other covert channels) can potentially be misused, we believe its advancement requires public in-depth studies for two reasons: (i) History covert channels have already shown to be applicable for censorship circumvention (see \textsc{OPPRESSION} \cite{AsiaCCS24:OPPR}), which aids the exchange of critical opinions through networks and, for this reason, supports an essential human right. (ii) When research results on network covert channels are made public, they aid the whole community, not just some of the actors (e.g., censors), allowing everybody to develop protection mechanisms. In contrast, when censors or cyber criminals develop such novel methods, they do not share their insights with the public, resulting in a disbalance of available knowledge.

\section{Conclusion}
We introduced a novel history covert channel based on network-timing pattern matching. Our approach offers significant improvements in stealthiness and robustness over previous methods, allowing for covert communication without modifying packet content. Our experimental evaluation demonstrates the practicality and effectiveness of our proposed method.

Robustness in the SHP algorithm is achieved through a combination of error tolerance, adaptive synchronization techniques, and intelligent use of network traffic patterns. By addressing the key challenges identified in the \ac{DYST} protocol, SHP improves its resistance to timing variability, noise, and detection.

To ensure continued robustness in increasingly hostile or dynamic network environments, further enhancements could be incorporated. Future research should focus on optimizing \ac{HCC}s for real-world deployment, including:
\begin{itemize}
    \item \textbf{Widening Input Sources:} Integrating a wider variety of hash input sources to improve data rate and improve robustness under diverse network conditions. 
    \item \textbf{Adaptive Pointer Strategies:} Implementing intelligent pointer selection mechanisms to dynamically adjust transmission parameters based on real-time network conditions and stabilize timing-based patterns.
    \item \textbf{Advanced Obfuscation Techniques:} Introduce different pointer techniques and randomized pointer delays to further reduce detectability.
    \item \textbf{Countermeasure Development:} Investigating effective defenses against \ac{HCC}s to improve network security policies and detection methods.
\end{itemize}

\bibliographystyle{elsarticle-num}
\bibliography{main}

\section*{Declarations}

\paragraph*{CRediT authorship contribution statement}

Christoph Weissenborn: Conceptualization, Methodology, Software, Validation, Formal analysis, Investigation, Data curation, Visualization, Writing – original draft. Steffen Wendzel: Supervision, Project administration, Conceptualization, Methodology, Validation, Writing – review \& editing.

\paragraph*{Declaration of competing interest}

The authors declare that they have no known competing financial interests or personal relationships that could have appeared to influence the work reported in this paper.

\paragraph*{Data and code availability}

All code and experimental data supporting the findings of this study are available for reproducibility and further research: \url{https://github.com/gXeeXqBHuHDFTaEnff3Z/SHP}.

\paragraph*{Funding}

This research received no external funding.

\paragraph*{Acknowledgments}

We thank Tobias Schmidbauer, Sebastian Zillien, and Jörg Keller for providing code and data of DYST~\cite{DYST22}, which we used for comparison. We also thank Haozhi Li, Tian Song and Yating Yang for their work on generic and sensitive anomaly detection \cite{GAS22}, which we used as approach for out ML-detection model.

\section*{Author Biographies}

\subsection*{}
\setlength\intextsep{0pt}
\begin{wrapfigure}{l}{0.16\textwidth}
\centering
\includegraphics[width=0.15\textwidth]{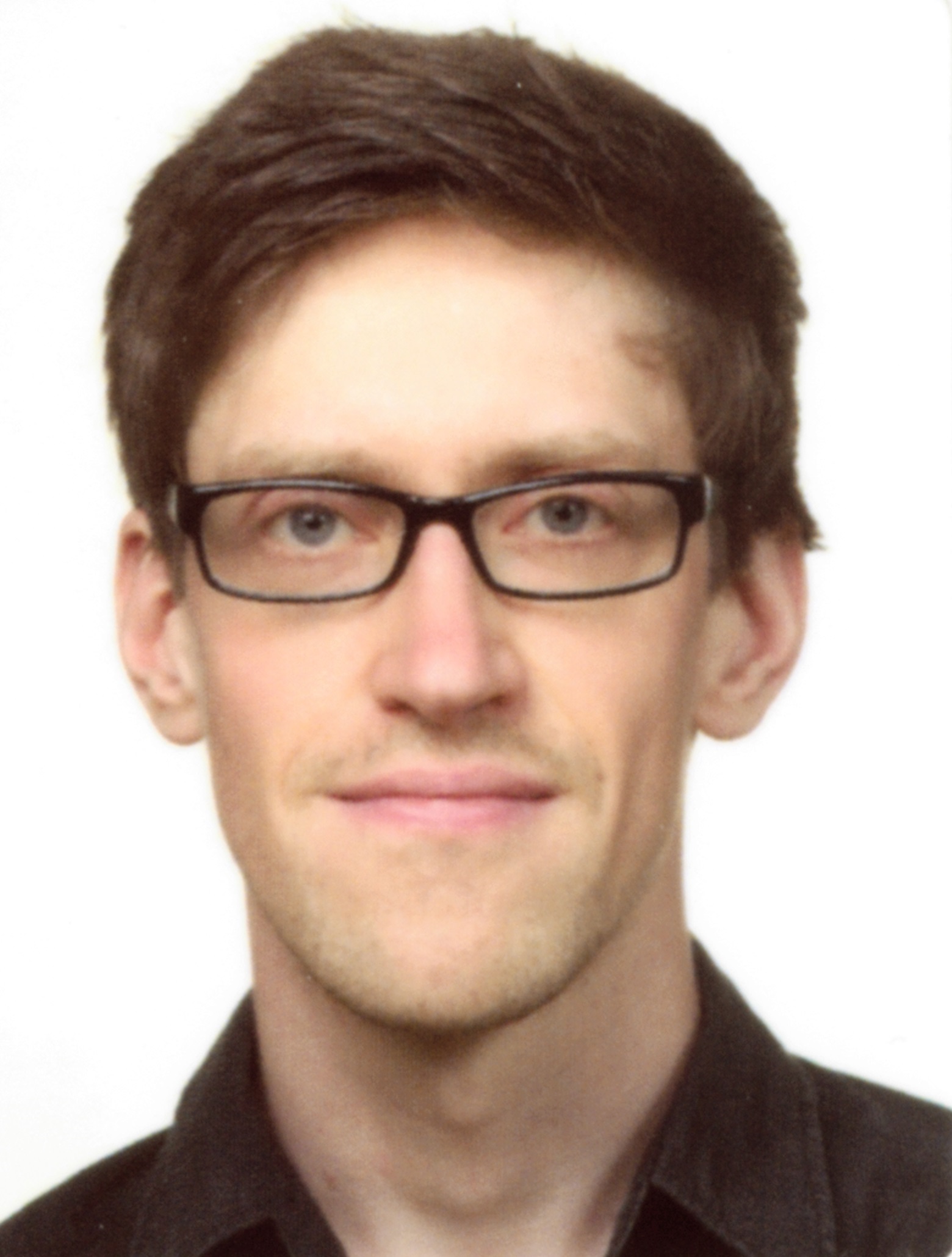}
\end{wrapfigure}
\noindent \textbf{Christoph Weißenborn} received his B.Sc. and M.Sc. degrees in Computer Science from FernUniversität Hagen in 2018 and 2022, respectively. Since 2024, he has been working at the German Federal Office for Information Security (BSI), where he leads the development of a standards framework for the automation of cybersecurity compliance. Previously, he worked at the German Federal Commissioner for Data Protection and Freedom of Information (BfDI), focusing on technical safeguards for telecommunications providers and data protection compliance. From 2019 to 2022, he was involved in national and international standardization efforts at BSI, participating in committees with ETSI, DIN, and ISO. His research interests include network security, cybersecurity frameworks and security standardization.

\subsection*{}
\setlength\intextsep{0pt}
\begin{wrapfigure}{l}{0.16\textwidth}
\centering
\includegraphics[width=0.15\textwidth]{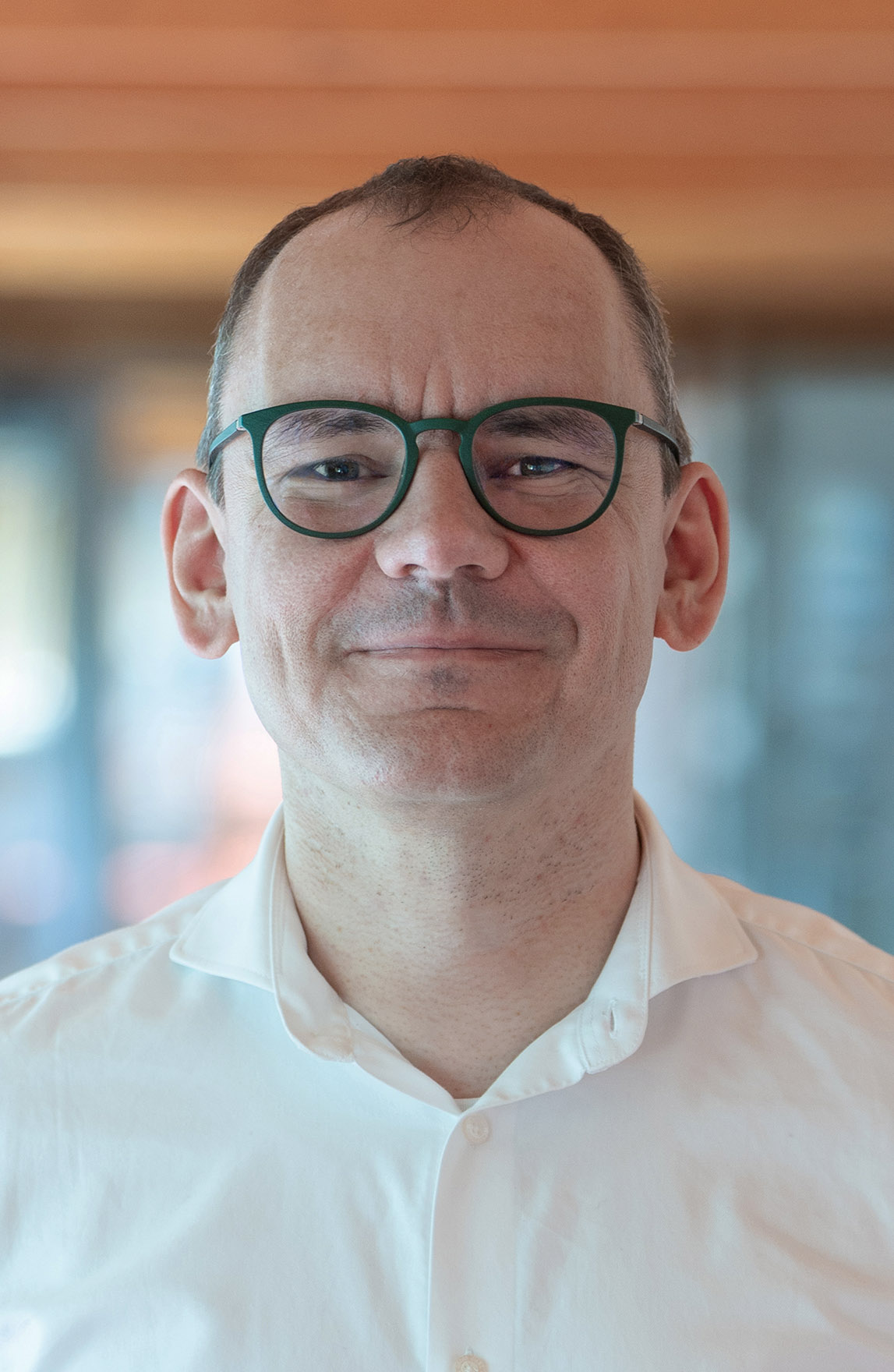}
\end{wrapfigure}
\noindent \textbf{Steffen Wendzel} is a professor at Ulm University, Germany, where he is the chair of the Institute of Information Resource Management (IRM) and the director of the university computing center and university library (kiz). Previously, he was a professor at HS Worms, Germany, a lecturer (\emph{Privatdozent}) at the Dep.\ Mathematics \& Computer Science, University of Hagen, Germany, and a PostDoc at Fraunhofer FKIE, Bonn, Germany. He received his PhD (2013) and Habilitation (2020) from the University of Hagen, Germany.
Steffen \mbox{(co-)}authored more than 190 publications, of which many appeared in major journals and conferences (e.g., Elsevier COSE, Elsevier FGCS, ACM CSUR, IEEE TDSC, IEEE S\&P Mag., Asia\-CCS, IEEE LCN, Comm.\ ACM, ARES etc.). Website: \url{https://www.wendzel.de} / \url{https://www.uni-ulm.de/en/in/omi/}.

\end{document}